\documentclass[]{interact}

\usepackage[natbibapa,nodoi]{apacite}
\renewcommand\bibliographytypesize{\fontsize{10}{12}\selectfont}
\usepackage{graphicx}
\usepackage{rotating}
\usepackage{enumerate}
\usepackage{amsfonts}
\usepackage{amsmath}
\usepackage{comment}
\usepackage{amssymb}
\usepackage{amscd,bm,amsbsy}
\usepackage{setspace}
\usepackage{amsthm}
\usepackage[framemethod=tikZ]{mdframed}

\DeclareMathOperator{\MA}{MA}
\DeclareMathOperator{\anc}{anc}

\DeclareMathOperator{\sib}{sib}
\DeclareMathOperator{\desc}{desc}

\newcommand{\bPhi}{\boldsymbol{\Phi}}

\newcommand{\bY}{\boldsymbol{Y}}

\newcommand{\bX}{\boldsymbol{X}}

\newcommand{\Name}{NSFS}

\mdfdefinestyle{mystyle2}{%
innertopmargin=10pt,%
innerbottommargin=10pt,%
innerleftmargin=10pt,%
innerrightmargin=15pt,%
roundcorner=15pt,%
}

\usepackage{epstopdf}
\usepackage[caption=false]{subfig}

\usepackage[longnamesfirst,sort]{natbib}
\bibpunct[, ]{(}{)}{;}{a}{,}{,}

\theoremstyle{plain}

\theoremstyle{definition}

\theoremstyle{remark}

\begin{document}

\articletype{Submitted to BOOM 2018}

\title{Analysis of Novel Annotations in the Gene Ontology for Boosting the Selection of Negative Examples}

\author{
\name{Maryam Sepehri\textsuperscript{a}\thanks{CONTACT Marco Frasca. Email: frasca@di.unimi.it} and Marco Frasca\textsuperscript{a}}
\affil{\textsuperscript{a} Computer Science Department, University of Milan, Via Comelico 39, Milan, Italy.}
}

\maketitle

\begin{abstract}
Public repositories for genome and proteome annotations, such as the Gene Ontology (GO), rarely stores negative annotations, i.e. proteins not possessing a given function. This leaves undefined or ill defined the set of negative examples, which is crucial for training the majority of machine learning methods inferring proteins functions.
Automated techniques to choose reliable negative proteins are thereby required to train accurate function prediction models.
 This study proposes the first extensive analysis of the temporal evolution of protein annotations in the GO repository. Novel annotations registered through the years have been analyzed to verify the presence of annotation patterns in the GO hierarchy. Our research supplied fundamental clues about proteins likely to be unreliable as negative examples, that we verified into a novel algorithm of our own construction, validated on two organisms in a genome wide fashion against approaches proposed to choose negative examples in the context of functional prediction.\\
\end{abstract}


\section{Introduction}
The Protein Function Prediction (PFP), which involves sophisticated computational techniques to accurately predict the annotations of new proteins and proteomes, is an emerging and challenging problem in the post-genomic era~\citep{Radivojac13}. An important issue in PFP is the selection of reliable negative examples for learning accurate predictors. The Gene Ontology (GO)~\citep{GO00}, the reference repository of protein functions, usually stores positive associations (annotations) between GO functions (also called terms) and gene products, whereas unannotated proteins are rarely marked as negative for a given term. If a protein is not currently annotated with a GO term, it could be either that the protein is a negative example for that term or a positive example which has not been detected yet due to insufficient investigations. The absence of a ``gold standard'' for negative examples calls for the design of techniques for the selection of negative examples to train accurate functional classifiers.

There has been relatively little study devoted to this issue, despite the enormous interest toward the prediction of protein functions. These approaches mostly relied upon the GO hierarchical structure to define the set of negative examples. 
The GO is structured as a directed acyclic graph (DAG) with different levels of specificity, where terms describing broader functions (e.g. DNA binding) are ancestors of those describing more specific functions (e.g. nucleosomal DNA binding). GO is composed of three branches, named  Biological Process (BP), Molecular Function (MF), and Cellular Component (CC). Annotations follows 
the \textit{true path rule} (TPR): a direct annotation of a protein with a given term must be transferred to all its ancestor terms in the hierarchy; thus, an annotation in nucleosomal DNA binding implies an annotation in DNA binding.

On this basis, earlier approaches considered as negative examples for a term all proteins with direct annotation (i.e., before applying the TPR) in neither descendant nor ancestral terms~\citep{Eisner05}, since proteins annotated with descendants are logically also member of the current term, and annotations (direct) with ancestors suggest just that future experiments might supply more specialized functions, thus annotating the protein to some of its descendants.
Proteins annotated with sibling terms (i.e. terms sharing at least one parent) have also been used as negative examples, under the assumption that proteins are rarely annotated in more than one child of the same parent term~\citep{Mostafavi09}. 
More recent works exploited the empirical conditional probability of annotating a protein with the GO function of interest given all the annotations of that protein with all the other functions~\citep{NOGO}, or just with the most specific ones~\citep{Youngs13}, in all three branches.
Lastly, in~\citep{Frasca17iwbbio} authors did not propose a novel strategy to select negatives, but assessed the relevance of some protein features extracted from protein networks in detecting false negatives using a GO temporal holdout setting.  

We propose here the first exhaustive study of annotation evolution through different releases of the GO, revealing novel interesting trends about existing and newly discovered protein functions.
Our research paid attention to the distribution of novel annotations in yeast and human organisms, and interestingly found that proteins tend to receive novel annotations for terms having high semantic similarity with the terms they were already annotated with. Thus, not just the hierarchical position is relevant in selecting reliable negatives, but even the term content plays a key role. 
The present study thereby provides a basilar knowledge for any future approach to detect negatives in the Gene Ontology.
Furthermore, by leveraging these emerging disclosures, we designed a novel method to select negative examples for GO terms, and assessed its effectiveness in an experimental comparison involving thousands of GO terms and the state-of-the-art methodologies proposed for the same task. 
\section{Preliminaries}\label{prelim}
Given a set $V=\{1,\ldots,n\}$ of proteins, which are annotated to $m$ GO terms $C = \{1, \ldots m\}$, by $\bY \in \{0,1\}^{n\times m}$ we denote the corresponding label matrix, where $Y_{ik} = 1$ if protein $i$ is annotated with term $k$, $Y_{ik} = 0$ otherwise.
Terms are organized in a DAG with different levels of specificity, where:
\begin{enumerate}[-]
\item The level of term $k$ in the hierarch is defined as  the number of edges on the maximum length path from a root node; 
\item $\anc(k) \subset \{1,\ldots,m\}$ is the set of ancestors of $k\in C$ in the hierarchy, that is all the terms on the paths from $k$ to the root term;
\item $\sib(k) \subset \{1,\ldots,m\}$ is the set of sibling terms of $k$, namely terms sharing at least one parent with $k$;
\item $\desc(k) \subset \{1,\ldots,m\}$ is the set of descendants of $k$, that is terms $s$ such that $k\in \anc(s)$;
\end{enumerate}
Forthermore, a map $\phi:C \times C \to \mathbb{R}$ is known, associating any pair of terms $k,r\in C$ with a similarity index $\phi(r,k)$. $\bPhi:= \phi_{rk}|_{k,r=1}^m$ is the resulting term similarity matrix, with $\phi(r,k) := \phi_{rk}$.

The \textit{temporal holdout} validation scheme relies on two different temporal GO releases, denoted by $\bY$ and $\overline\bY$, assuming $\bY$ as the
older one. Columns $\bY_{.k}$ and $\overline\bY_{.k}$ represent thereby the labels/annotations for term $k$ in the older and later release, respectively. Here $\bX_{.r}$ and $\bX_{i.}$ denote the $r$th column and the $i$th row of a matrix $\bX$, respectively.
Moreover, fixed a term $k$, $V^{k}_{np}\subset V$ is the set of proteins that received novel 
annotations for $k$ in the holdout period, i.e. $V_{np}^{k} = \{i\in V | Y_{ik}=0 \wedge \overline Y_{ik}=1 \}$. When clear from the context, we denote $V^{k}_{np}$ simply by $V_{np}$.

The \textit{negative selection} problem consists in learning a model to accurately discriminate proteins belonging to $V_{np}$  from proteins $\{i\in V | Y_{ik}=0 \wedge \overline Y_{ik}=0 \}$ (i.e. not annotated in both releases).

\section{Data}
Functional annotations have been downloaded from the Gene Ontology for the \textit{S.cerevisiae} (yeast) and \textit{Homo sapiens} (human) organisms.
Three different temporal releases have been considered: the UniProt GOA releases $69$ (9
May 2017) as novel release, and releases $52$ (December 2015) or $40$ (November 2014) in turn as older release for \textit{yeast},  and releases $168$ (May 2017), $151$ (December 2015) or
$139$ (November 2014) for \textit{human}. Annotations include all available terms in the three GO branches. 
In order not to introduce label biases, in both releases solely experimentally validated annotations have been considered.
The number of proteins in the set $V_{np}$ and the total number novel annotations in the holdout period are reported in Table~\ref{tab:novel_ann}. The analysis only considers the direct annotations.
\begin{table}[!t]
\scriptsize
\centering
\begin{tabular}{lcccc}
  \hline
 {\tt GO branch} & {\tt Proteins} & {\tt Novel annotations} & {\tt Proteins} & {\tt Novel annotations}\\ 
  \hline\\[-5pt]
& \multicolumn{2}{c}{\textbf {yeast}} & \multicolumn{2}{c}{\textbf {human}}\\[5pt]
&  \multicolumn{2}{c}{\textbf{ GO releases 69, 52}} & \multicolumn{2}{c}{\textbf { GO releases 168, 151}}\\[3pt] 
CC & 167 & 217 & 3878 & 9406 \\[2pt] 
  MF & 184 & 261  & 2012 & 5059\\[2pt] 
  BP & 336 & 586  & 1460 & 10769 \\[5pt] 
&  \multicolumn{2}{c}{\textbf{ 69, 40}} &  \multicolumn{2}{c}{{\textbf {168, 139}}}\\[5pt] 
CC & 319 & 437 & 4545 & 13230\\[2pt] 
  MF & 381 & 565  & 4483 & 9970\\[2pt] 
  BP & 601 & 1097 & 2377 & 19120 \\[5pt]
  \hline      
   \hline\\
\end{tabular}
\caption{Number of proteins received novel annotations in the most recent release and the corresponding total number of novel annotations.}\label{tab:novel_ann} 
\normalsize
\end{table}
\section{Analysis of novel annotations}\label{sec:analy}
We propose in this section an analysis of the distribution of proteins $V_{np}$ across the GO hierarchy, with a twofold aim: verify whether novel annotations show any trend or pattern in the hierarchy; verify the validity of hypotheses existing approaches relying on the GO structure to select negatives build upon.
\subsection{Locating novel annotations in the hierarchy} \label{sub:locating} 
Given $i\in V_{np}^k$, let $C_i\subset C$  be the subset of GO terms the protein $i$ is annotated in the older release, that is $C_i = \{r \in C | Y_{ir} = 1\}$ ($k\notin C_i$ by definition of $V_{np}^k$).  
First of all, it might happen that $C_i = \emptyset$, i.e. the protein $i$ had no previous annotations (this case is denoted by \textit{First}). Otherwise, given $s \in C_i$, the following cases are distinguished:
\begin{list}{}{}
\item \textit{Anc}: $k \in \anc(s)$,
\item \textit{Desc}: $k \in \desc(s)$,
\item \textit{Sib}: $k \in \sib(s)$,
\item \textit{Other}: $k \notin \{\anc(s)\cup\desc(s)\cup\sib(s)\}$.
\end{list}

\begin{figure}[t]
\begin{mdframed}[style=mystyle2]
\begin{center}
\begin{tabular}{cc}
\hspace{-0.5cm} (a)  & \hspace{-0.3cm} (b)\\[-7pt]
\hspace{-0.5cm} \includegraphics [angle=-90,width=0.5\textwidth] {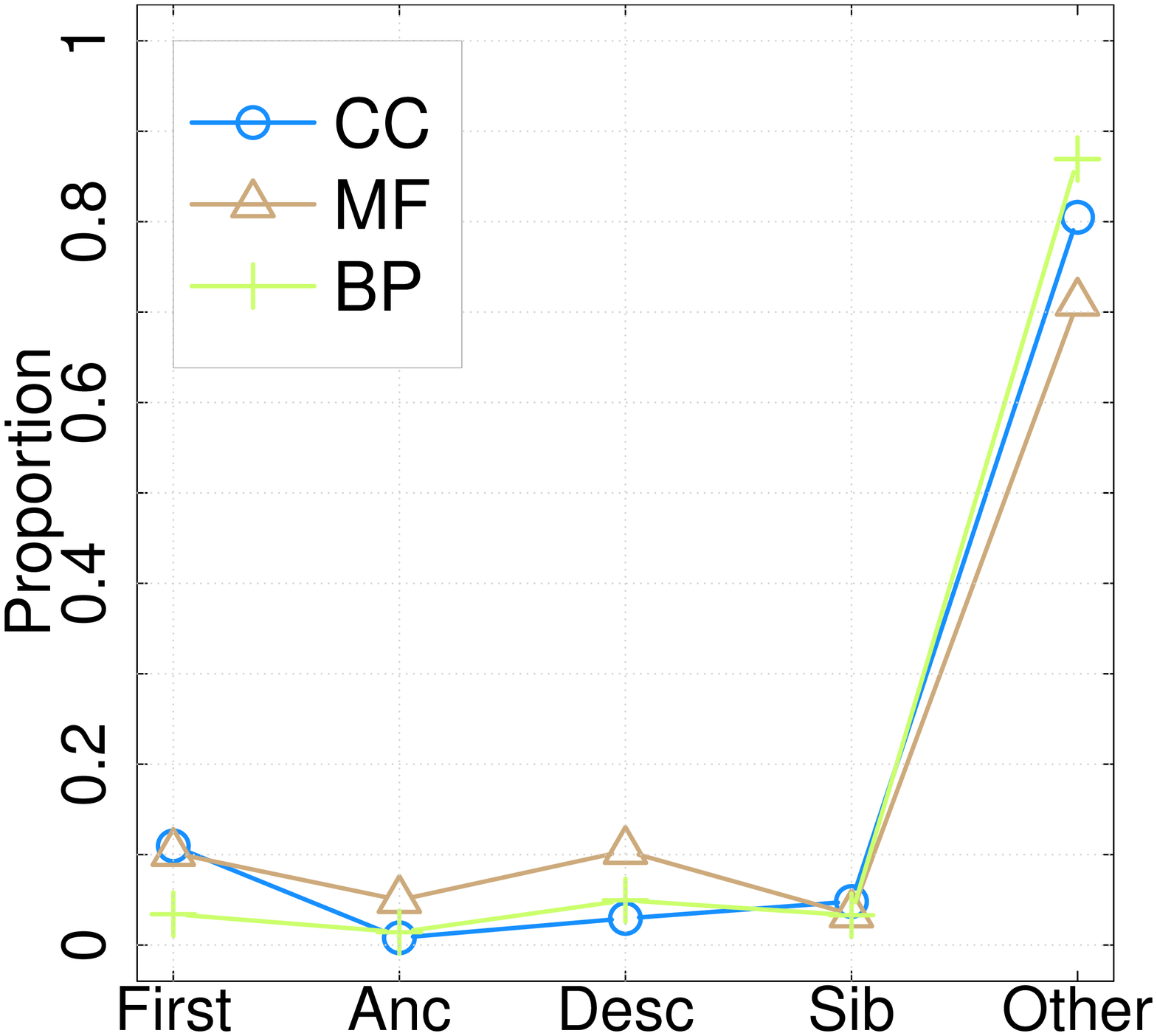}& 
\hspace{-0.3cm} \includegraphics [angle=-90,width=0.5\textwidth] {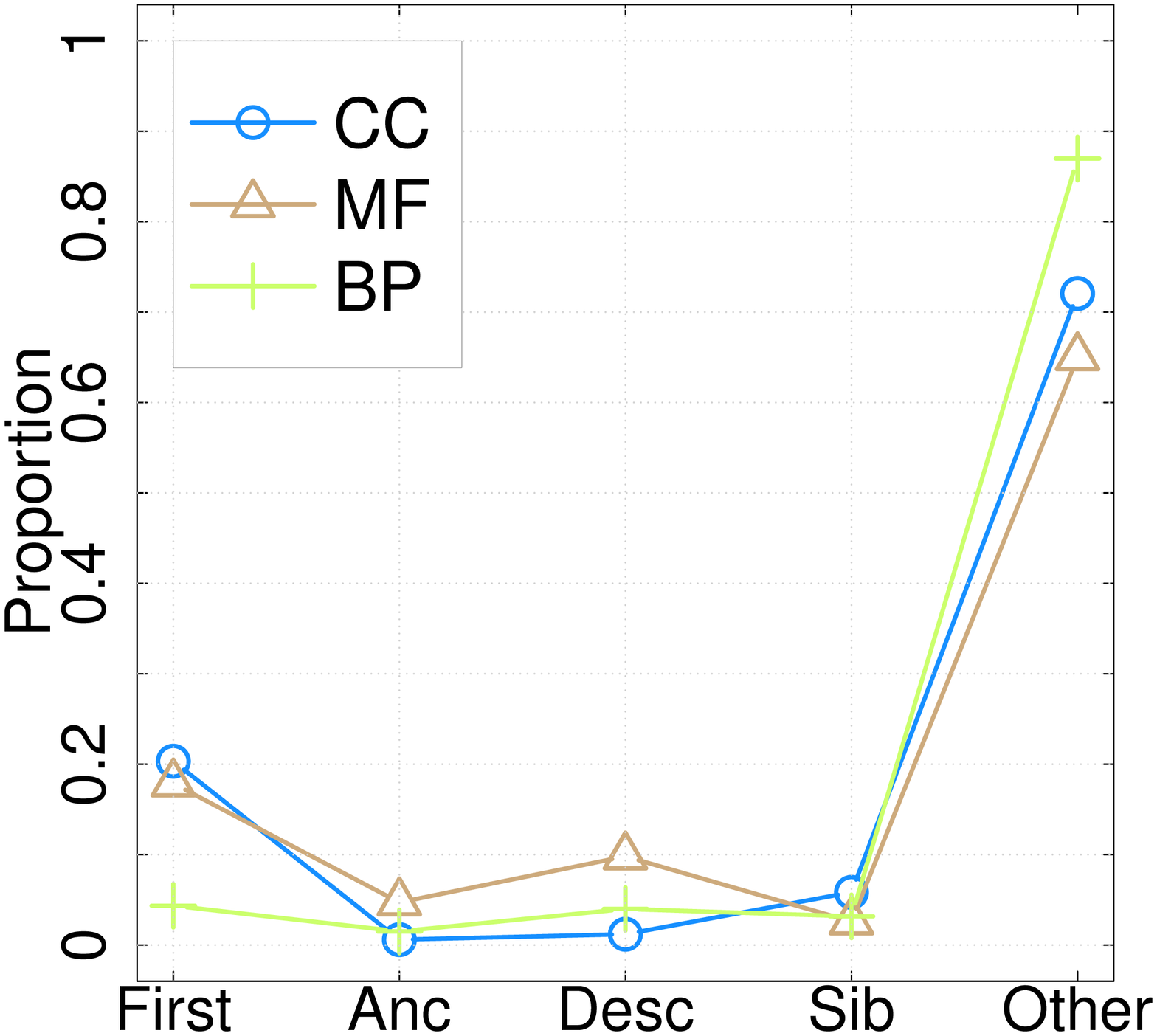}\\\\[0pt]
 \hspace{-0.5cm} (c)  & \hspace{-0.3cm} (d)\\[-7pt]
\hspace{-0.5cm} \includegraphics [angle=-90,width=0.5\textwidth] {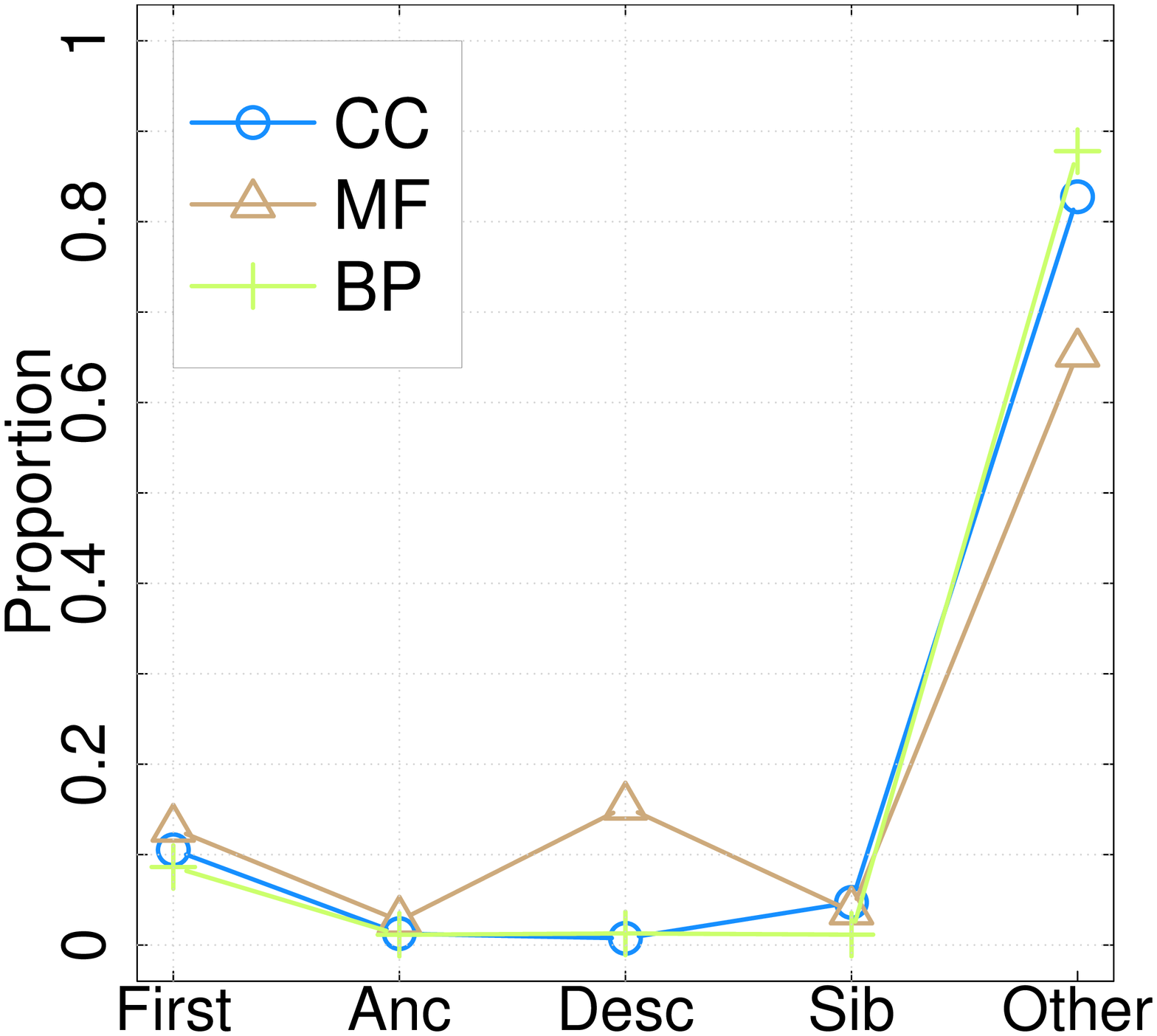}& 
\hspace{-0.3cm} \includegraphics [angle=-90,width=0.5\textwidth] {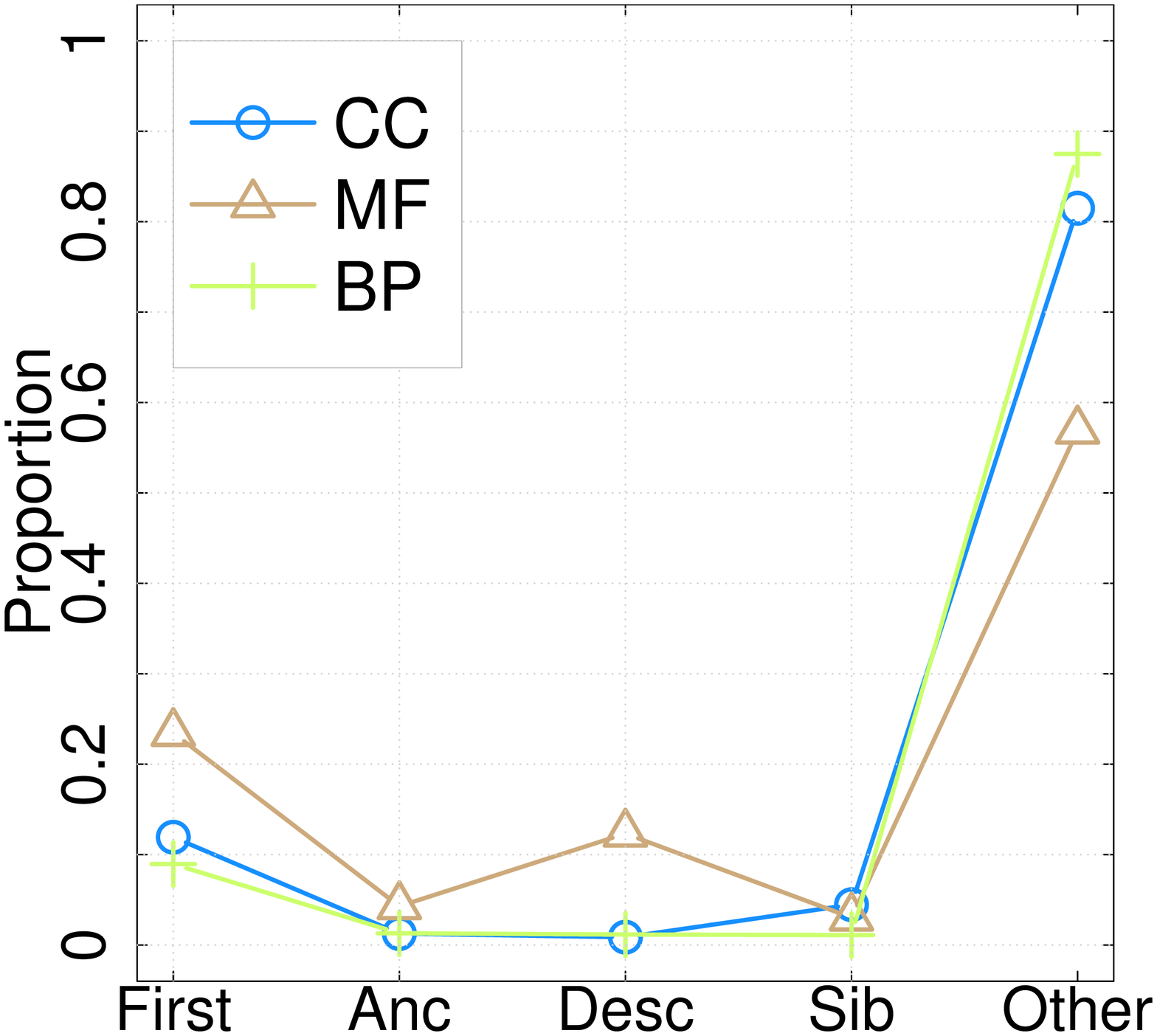}\\
\end{tabular}
\end{center}
\caption{Proportion of proteins $V_{np}^k$ falling in one of the five considered cases averaged across terms $k$ for yeast using UniProt GOA releases (a) $69$ / $52$  and (b) $69$ / $40$  and Human using UniProt GOA releases (c) $168$ - $151$  and (d) $168$ - $139$. \label{fig:3dSim}}
\end{mdframed}
\end{figure}

Figure~\ref{fig:3dSim} depicts the obtained distribution of novel annotations, considering solely direct annotations for both yeast and human organisms.  
First of all, around $5\%$ of novel annotations are for sibling terms: for instance yeast protein ADY3 (protein required for spore wall formation) had an annotation for term GO:0030476 (ascospore wall assembly) on 2002/09/20 -- 	PMID:11973299, and on 2016/06/29 has been annotated with  GO:0032120 (ascospore-type prospore membrane assembly, PMID:11742972), sibling of GO:0030476 via GO:1903046 (meiotic cell cycle process). Thus, strategies adopting as negatives the proteins positive for sibling terms in these cases fail. Another relevant problem  of the sibling approach is that for several terms (those with few annotations available) it does not find any negative example (no annotations available for sibling terms). 

Non zero is also the proportion associated with ancestor terms, especially for the MF ontology, yeast data: this is not surprising, since a protein can receive a direct annotation for an ancestor term, even when it is already annotated for a descendant term (remind the figure refers to non-TPR annotations). For instance, the yeast protein KAR9 (Karyogamy protein KAR9) was annotated with GO term GO:0005938 (cell cortex) on 2002/12/02 (PMID:9442113), and later (2016/04/22 -- PMID:26906737) has been annotated also with term GO:0005737 (cytoplasm), which is ancestor of GO:0005938.
A relevant number of novel annotations is for descendant terms, especially for human data and MF ontology. This is quite obvious, since novel studies tend to reveal more specific roles and functions of proteins. However, still around $80\%$ of cases do not fall into neither  \textit{Anc} nor \textit{Desc} categories, suggesting that strategies selecting negatives  among proteins with direct annotations in neither descendant nor ancestral terms might incur into a considerable number of false negatives. This hint is experimentally verified in Section~\ref{sec:negsel}.

First annotations for yeast proteins are more frequent in the CC and MF branches on yeast data, likely due to the higher complexity of biological processes, making more difficult finding novel annotations. Finally, BP terms have the highest proportions  in the category `other', which includes all remaining cases.

Since the category `other' collect most of novel annotations, we carried on other two experiments to understand more about them, one exploiting the semantic similarities between GO terms, another based on the GO hierarchical structure.
\subsection{Semantic similarity analysis}
Recalling $C_i\subset C$ is the set of GO terms the protein $i\in V_{np}^k$ was already annotated in the older release, and that by definition $k\notin C_i$, we studied the semantic similarities between $k$ and terms in $C_i$, to detect eventual trends in their distribution. The semantic similarities of term $k$ are contained in the vector $\bPhi_{k.}$ (see Section~\ref{prelim}), and for each term $s\in C_i$ we are interested in the rank $r_{ks}$ of $\phi_{ks}$ within the vector $\bPhi_{k.}$: $r_{ks}:= q$ if $\phi_{ks}$ is the $q$-th largest value in $\bPhi_{k.}$, or in other words  $\phi_{ks}$ occupies position $q$ in the vector $\bPhi_{k.}$ sorted in increasing way. We then normalize the rank dividing it by $m$. 

 As map $\phi$ we considered two state-of-the-art similarity measures, the \textit{Lin}~\citep{Lin98} and \textit{Jaccard}~\citep{Jaccard02} measures. 
Denoted by $\nu(k)$ the frequency of proteins annotated with term $k$, and by $\MA(k,r)$ the common ancestor of terms $k$ and $r$ whose frequency $\nu(\MA(k,r))$ is the lowest among all ancestors of $k$ and $r$, the Lin similarity between terms $k$ and $r$ is defined as follows: 
\begin{displaymath}
\phi_1(k,r) = \frac{2\log\nu(\MA(k,r))}{\log \nu(k) + \log \nu(r)}~.
\end{displaymath}
The value $-\log(\nu(k))$ in Information Theory is information content of term $k$, so that the higher the frequency, the lower the information carried by the term. Thus, the Lin similarity depends on the ratio between the information content of the least common ancestor and the information content of $k$ and $r$. Observing that by definition $\MA(k,r) \ge \max\{\nu(k),\nu(r)\}$, it follows that $0\leq \phi_1(k,r) \leq 1$.

Let $V_+^{k}$ be the set of proteins annotated with term $k$, the Jaccard similarity measure of terms $k$ and $r$ is:
\begin{displaymath}
\phi_2(k,r) =  \left\{ \begin{array}{cl}
{\displaystyle \frac{\big|V_+^{k}\cap V_+^{r}\big|}{\big|V_+^{k}\cup V_+^{r}\big|} } & \text{if $V_+^{k}\cup V_+^{r} \neq \emptyset$} \\
\\
0  & \text{otherwise.}
\end{array}
\right.
\end{displaymath}
This is the ratio between the number of proteins annotated with both terms and the number of proteins annotated with at least one term. The higher the number of shared annotations, the higher the similarity (up to $1$). When two terms do not share any positive example, their similarity is zero. In a hierarchy of terms like GO, terms with many annotations are usually closer to the root (less specific). In this case the denominator of $\phi_2$ tends to reduce the similarity between the two terms as opposed to the case in which terms have few annotations. Indeed, sharing annotations between two specific terms (closer to leaves) is more informative than sharing annotations between two more general terms.

\begin{figure}[t]
\begin{mdframed}[style=mystyle2]
\begin{center}
\small
\begin{tabular}{ccc}
\hspace{-0.35cm} (a) CC & \hspace{-0.6cm} (b) MF & \hspace{-0.6cm} (c) BP\\[-5pt]
\hspace{-0.35cm} \includegraphics [angle=-0,width=0.3\textwidth] {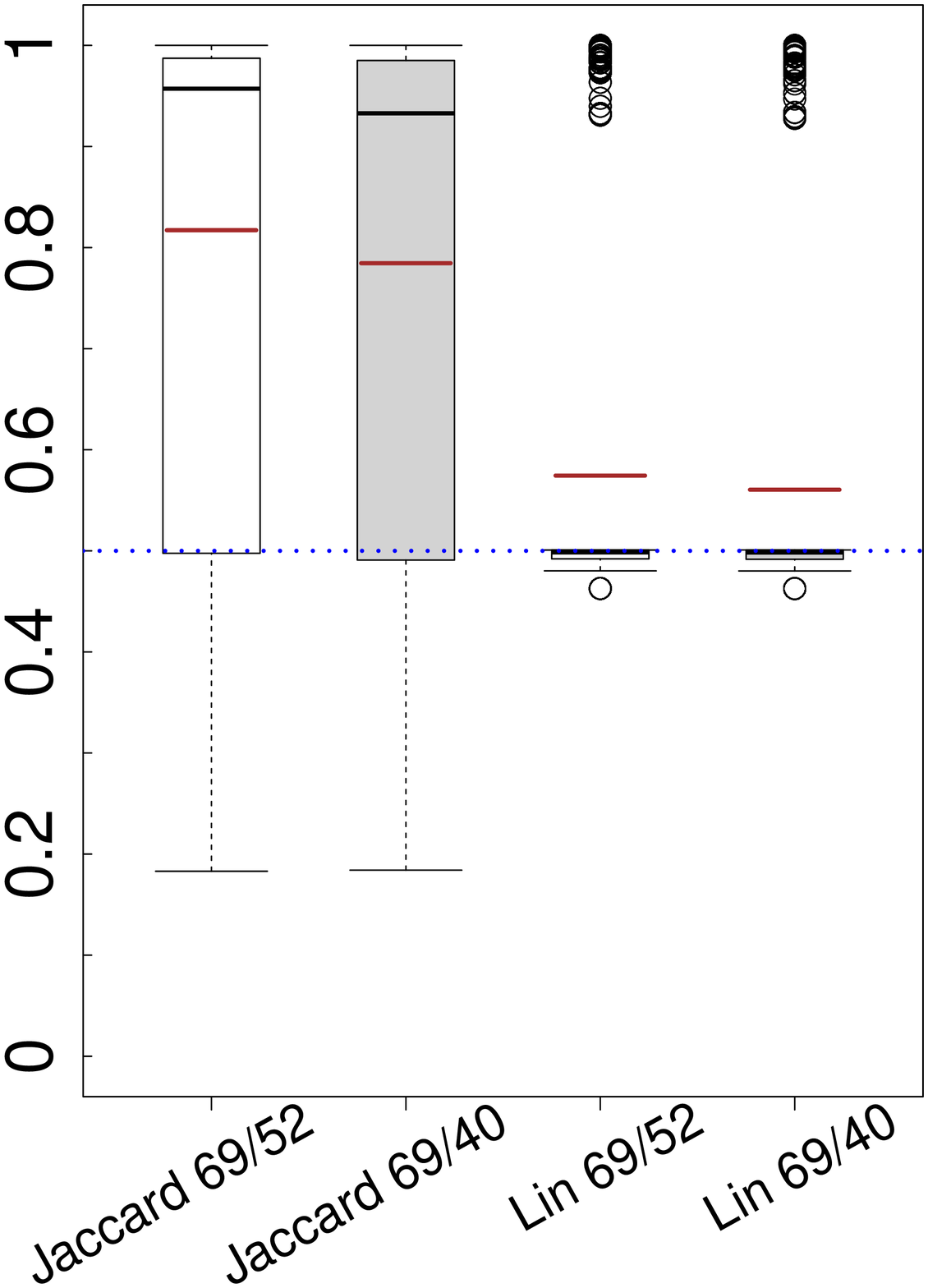}& 
\hspace{-0.35cm} \includegraphics [angle=-0,width=0.3\textwidth] {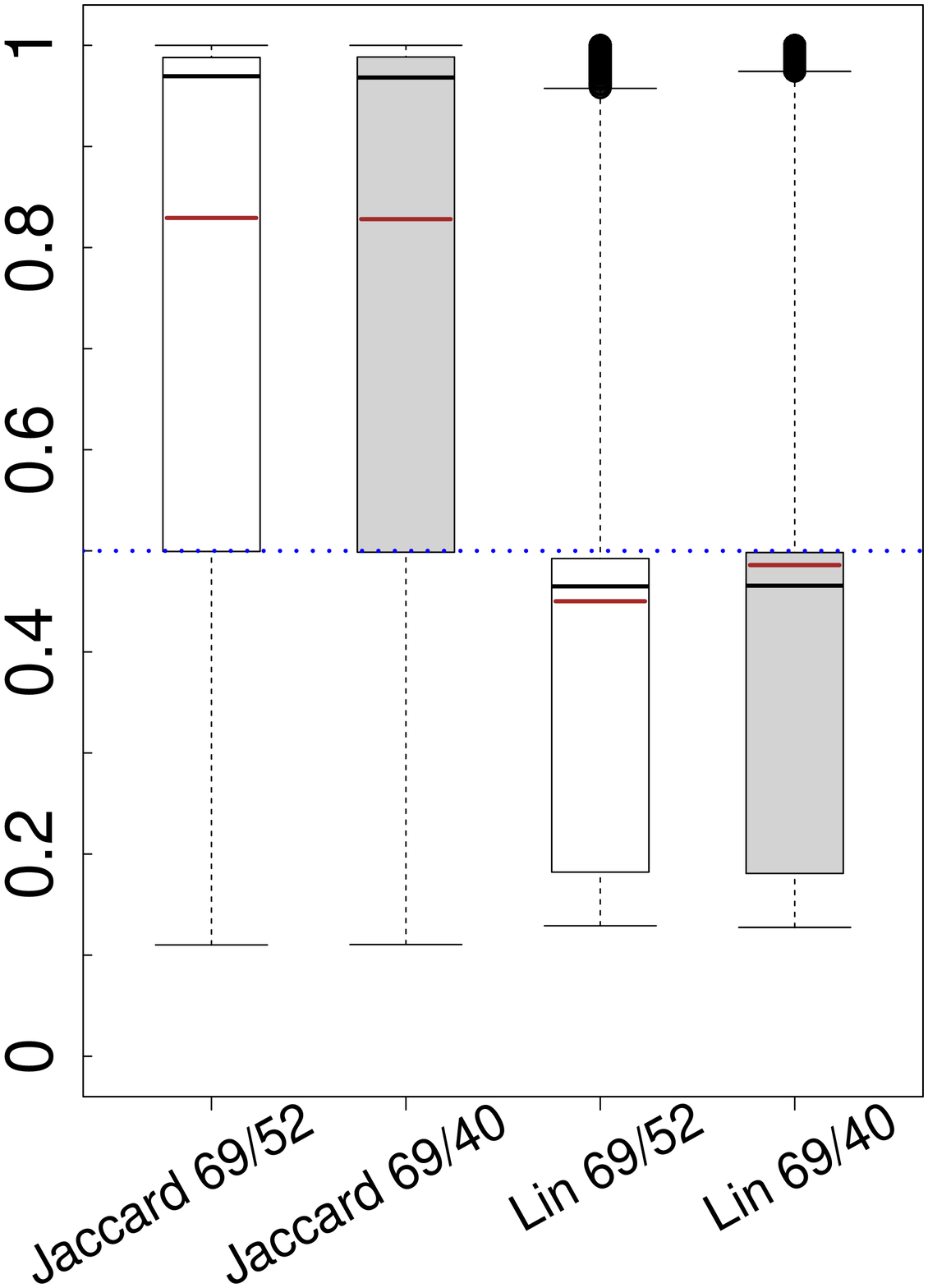}&
\hspace{-0.35cm} \includegraphics [angle=-0,width=0.3\textwidth] {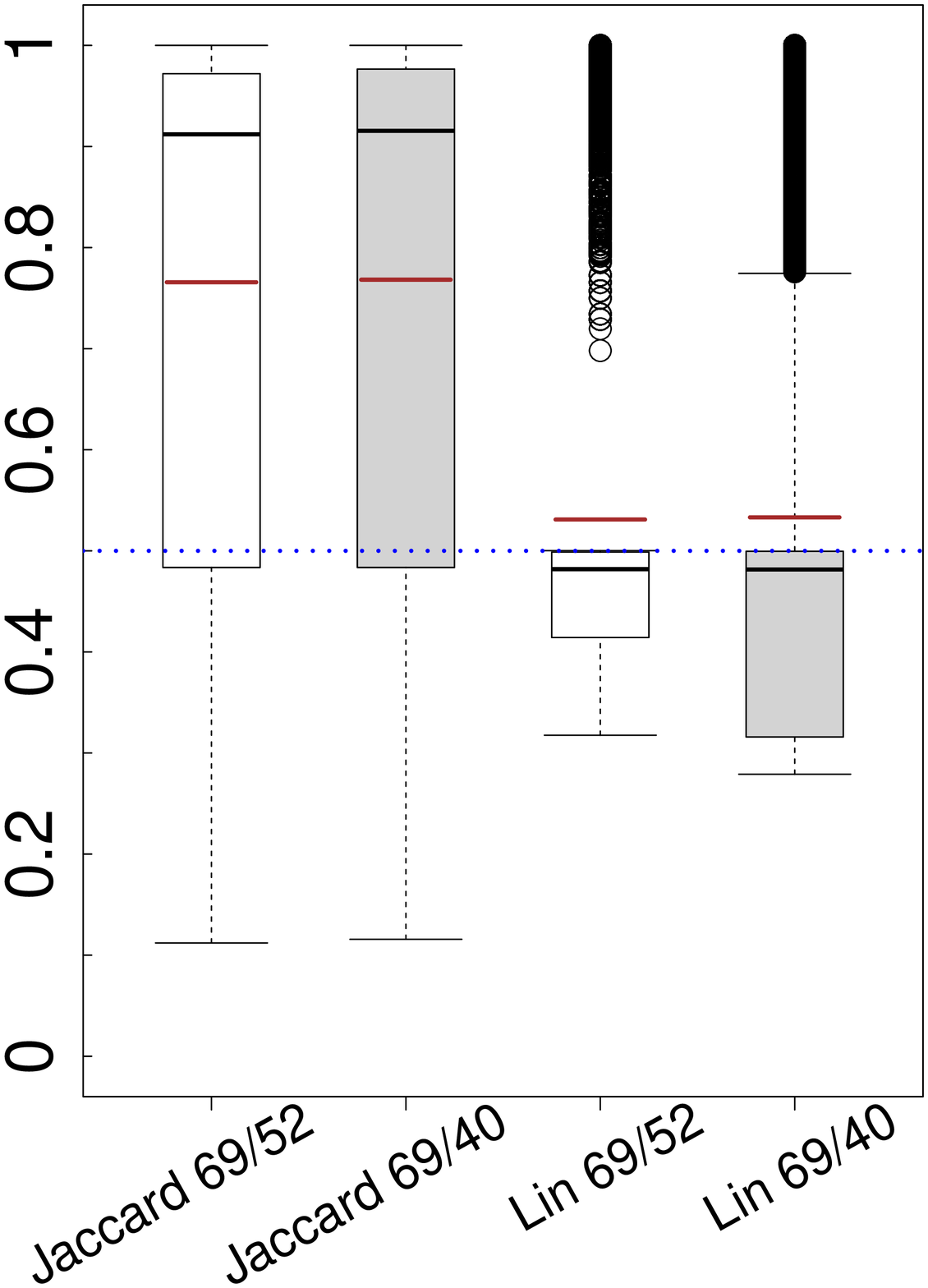}\\[5pt]
\hspace{-0.35cm} (d) CC & \hspace{-0.6cm} (e) MF & \hspace{-0.6cm} (f) BP\\[-5pt]
\hspace{-0.35cm} \includegraphics [angle=-0,width=0.3\textwidth] {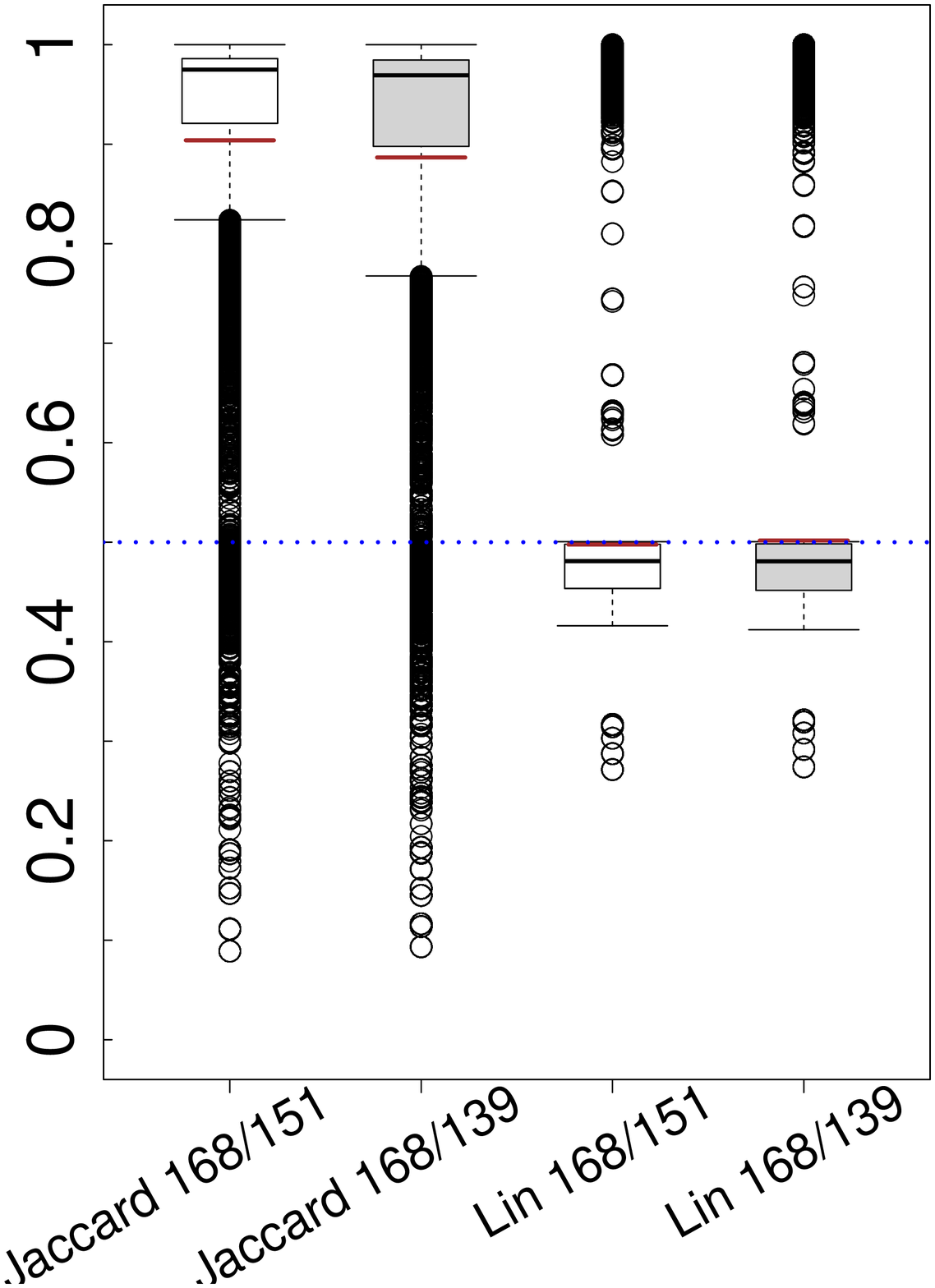}& 
\hspace{-0.35cm} \includegraphics [angle=-0,width=0.3\textwidth] {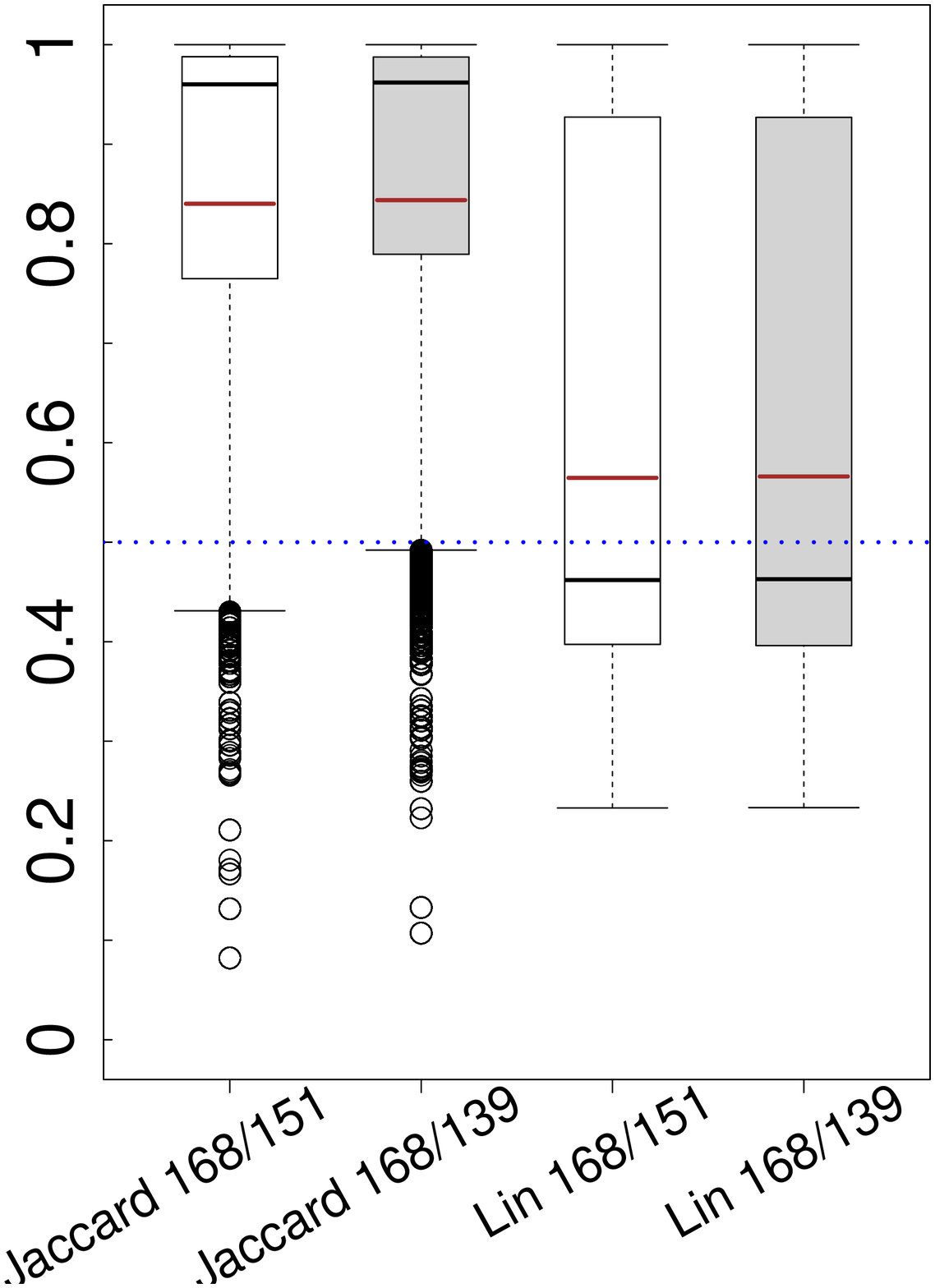}&
\hspace{-0.35cm} \includegraphics [angle=-0,width=0.3\textwidth] {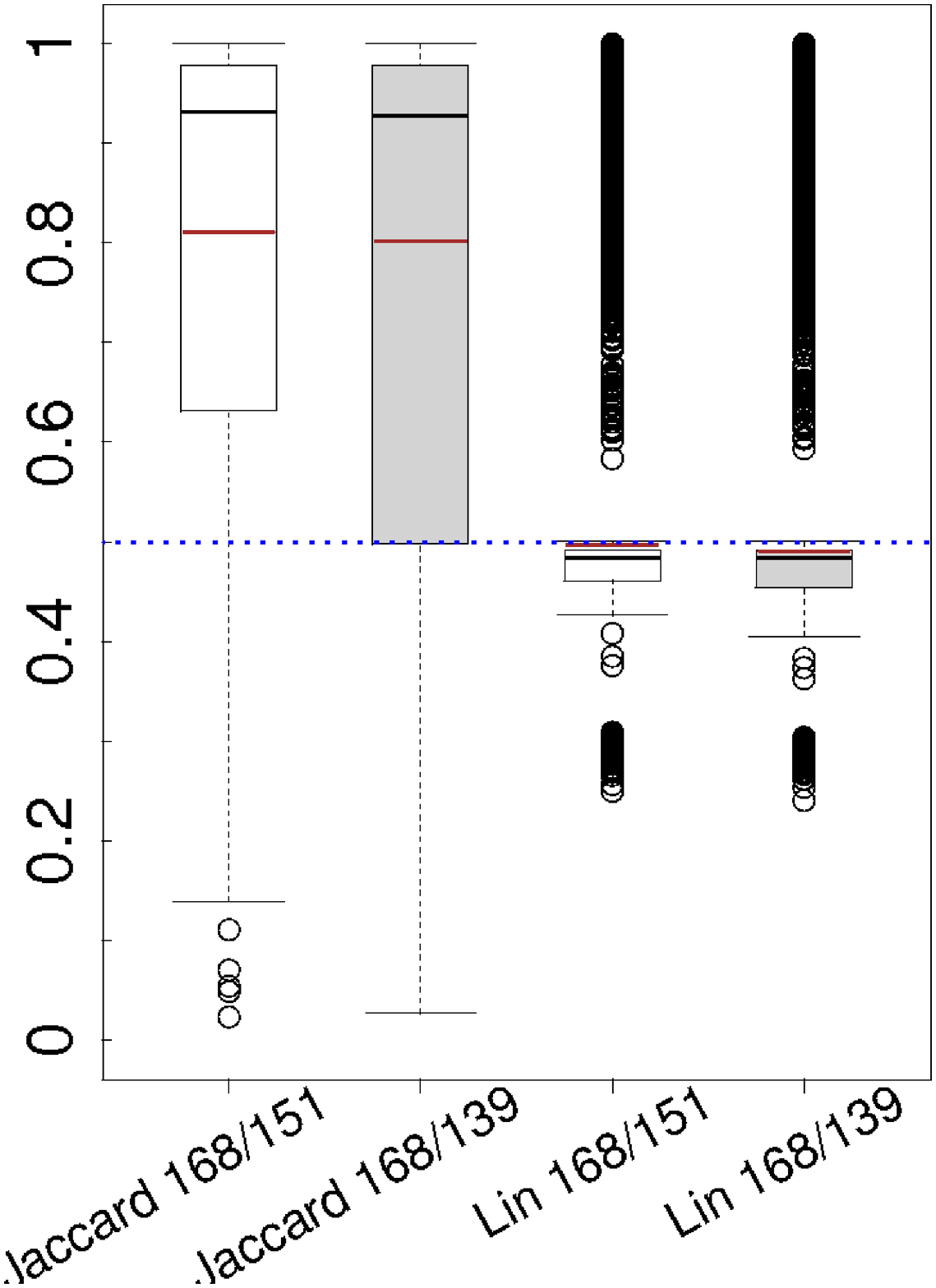}
\end{tabular}
\end{center}
\caption{Empirical distribution of term similarity rankings between the terms $C_i$ and term $k$, for proteins $i\in V_{np}^k$. 
Solely direct annotations are considered. Figures (a), (b) and (c)  and Figures (d), (e), (f) correspond to \textit{yeast} and \textit{human} data,  respectively.\label{fig:GO_rankings}}
\normalsize
\end{mdframed}
\end{figure}

Figure~\ref{fig:GO_rankings}  depicts the boxplots of the obtained ranks for all proteins $V_{np}^k$, averaged across terms $k$, considering both Lin and Jaccard similarity measures. As further information, the rank means are also drown as red horizontal segments. Lin and Jaccard measures led to much different rankings in all ontology and organisms. Indeed, Lin ranks are often below $0.5$ in 3/4 of cases, which does not provide clear clues about future annotations of protein $i$.
Moreover, rank means in this case are often higher than medians, meaning that the ranks are distributed toward lower values, with some high rank outliers. 
As opposite, according to the Jaccard similarity at least three terms out of four have a rank higher than $0.5$: in all experiments  more than half of terms in $C_i$ are in  among the $10\%$ of terms most similar to term $k$. On human data   (Figure~\ref{fig:GO_rankings} (d), (e), (f)) such trend is still more marked. In other words, given a protein $i$ annotated for a set of terms $C_i$ in the older release, in average every novel annotation of $i$ in the later release is for terms $k$ such that at least half of terms $C_i$ results in the top $10\%$ terms most similar to $k$. As further confirm, rank means in this case are always lower than medians, denoting the presence of some outliers with low rank, a large minority of  terms $C_i$, whereas remaining terms tend to be very similar to $k$. We propose a strategy for selecting negatives exploiting this results in Section~\ref{sec:negsel}.
\subsection{Distance from the closest path split}
Given $i\in V_{np}^k$, we also investigated how the term $k$ is located with respect to terms $C_i$. In particular, for any term $s\in C_i$, the novel annotation for $i$ with term $k$ falls in one of the following cases:
\begin{enumerate}[1.]
\item $k \in \anc(s)$. No novel annotation path is generated;
\item  $k \in \desc(s)$. The previous path from the root to $s$ is extended till $k$;
\item $k \in \sib(s)$. A new path fork is generated from a common parent term of $k$ and $s$ to term $k$;
\item $k \notin \{\anc(s)\cup\desc(s)\cup\sib(s)\}$. In this case, there is a common ancestor $q\in \anc(k)\cap\anc(s)$ where a new annotation path fork is created, namely from  $q$ to $k$. For consistency, we assume the three branch roots have a parent `dummy' node assumed as root of the DAG.
\end{enumerate}
By annotation path we intend here a path (sequence of  connected nodes) in the GO DAG such that protein $i$ is annotated with each node on the path (with TPR functional transfer). 
Since GO terms are specializations of their parents, the distance on a given path between GO terms is related to a semantic relatedness of the corresponding functions. Accordingly, if the distance from $q$ to $s$ (intended as the number of edges on the longest path between $q$ and $s$) is large, terms $s$ and $k$ tend to correspond to more different concepts; on the contrary, when $q$ and $s$ are closer, $s$ and $k$ are expected to describe concepts more semantically related.
We adopted the maximum distance between $s$ and $q$ in order to consider the  worst case results. Furthermore, since multiple paths in a DAG might be available from the root to a given node, and in principle there could be more than one $q\in \anc(k)\cap\anc(s)$ generating a new annotation path fork, we chose in this case that with the highest level -- thus more specific, see Section~\ref{prelim}.   

\begin{figure}[t]
\begin{mdframed}[style=mystyle2]
\begin{center}
\small
\begin{tabular}{ccc}
\hspace{-0.6cm} (a) CC & \hspace{-0.5cm} (b) MF & \hspace{-0.3cm} (c) BP\\[-2pt]
\hspace{-0.6cm} \includegraphics [angle=-0,width=0.3\textwidth] {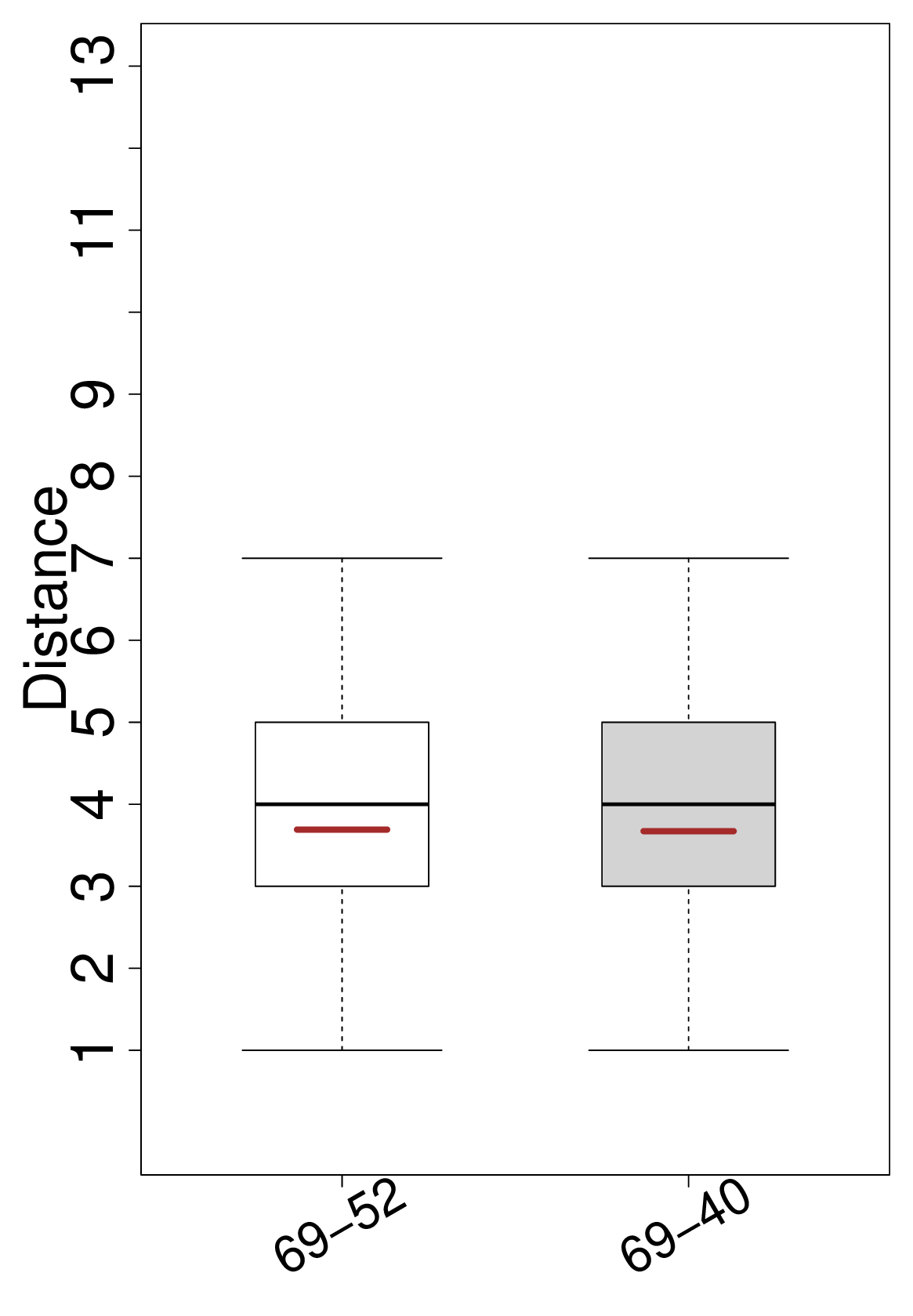}& 
\hspace{-0.5cm} \includegraphics [angle=-0,width=0.3\textwidth] {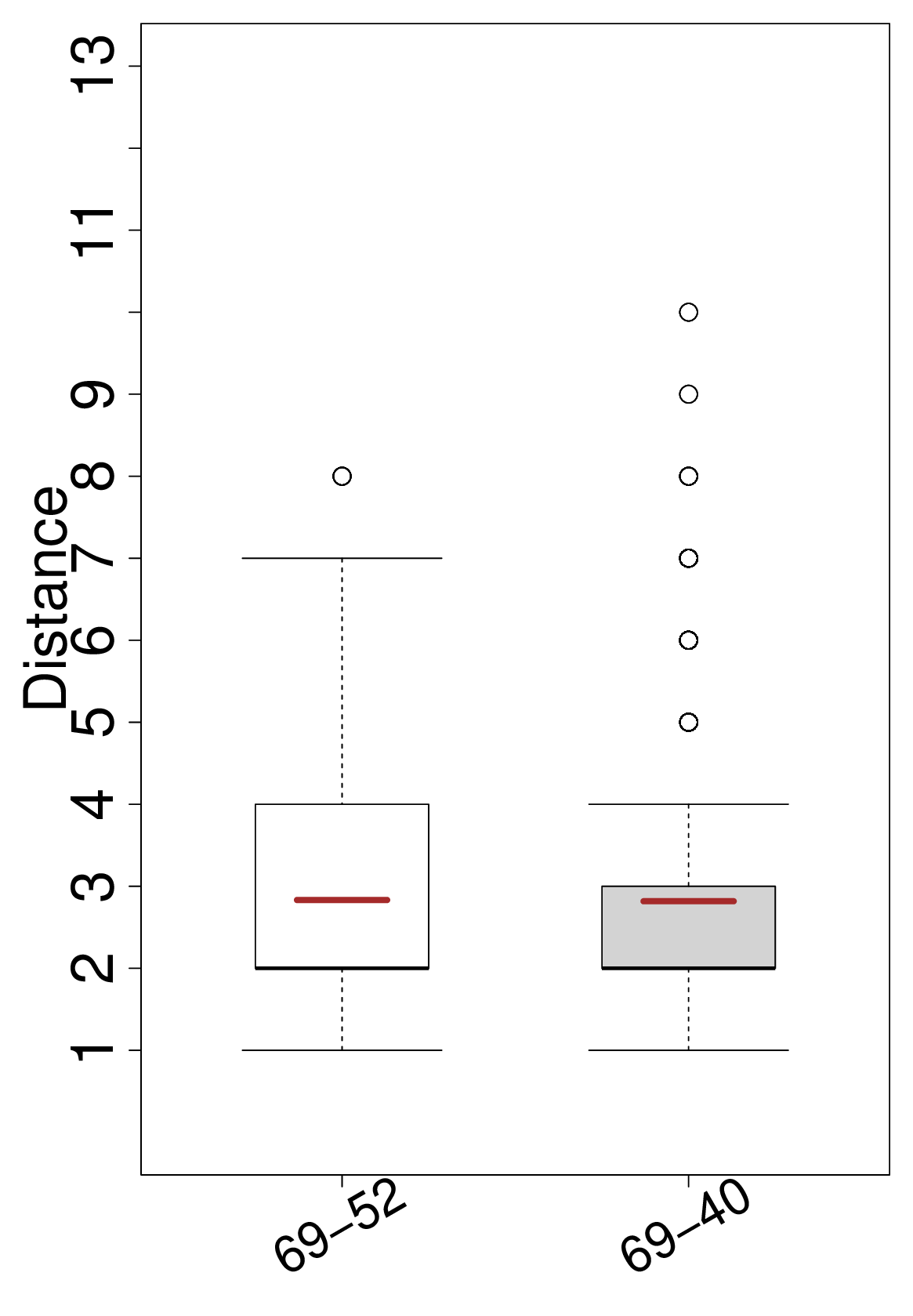}&
\hspace{-0.55cm} \includegraphics [angle=-0,width=0.3\textwidth] {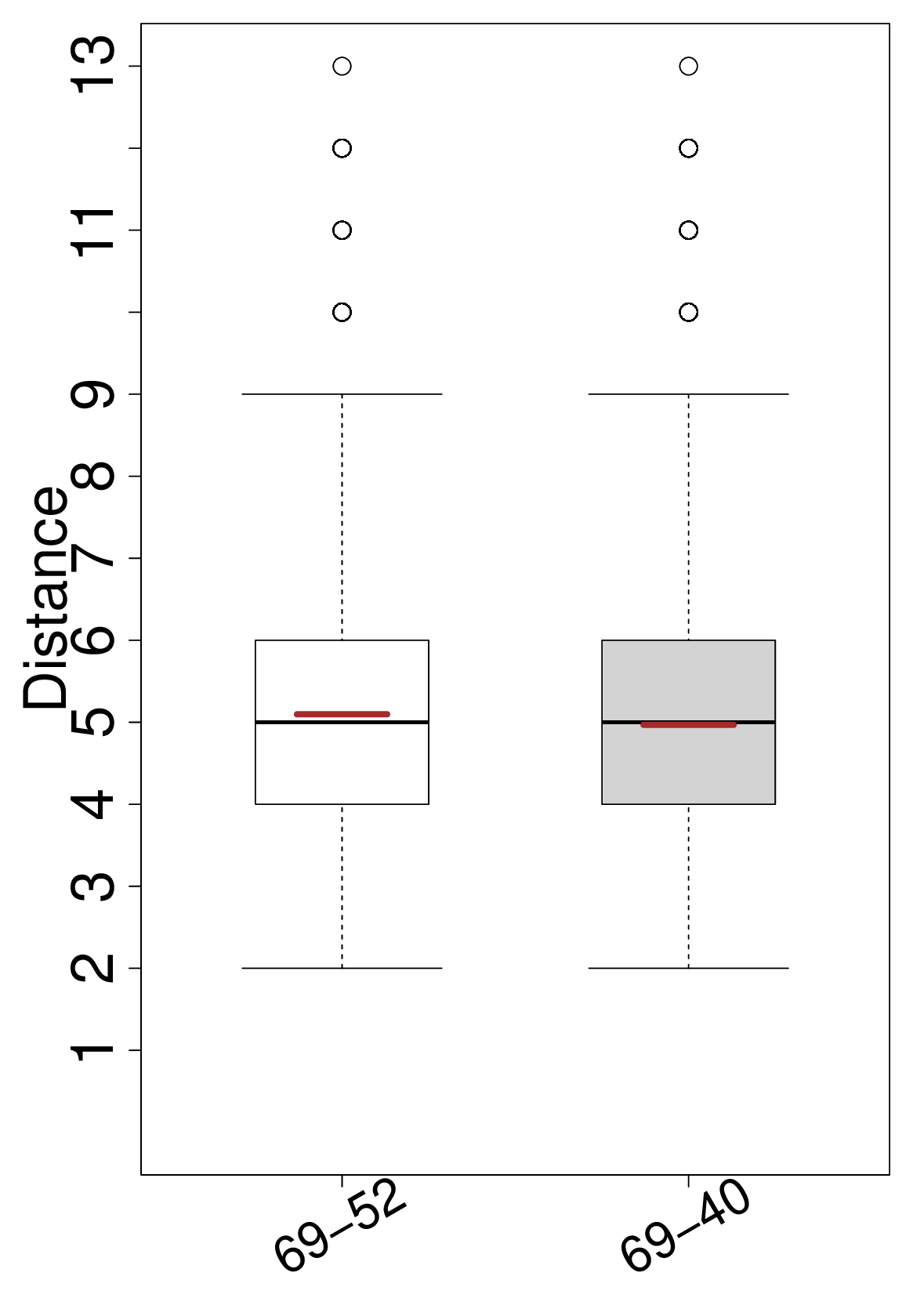}\\[5pt]
\hspace{-0.6cm} (d) CC & \hspace{-0.3cm} (e) MF & \hspace{-0.3cm} (f) BP\\[-2pt]
\hspace{-0.6cm} \includegraphics [angle=-0,width=0.3\textwidth] {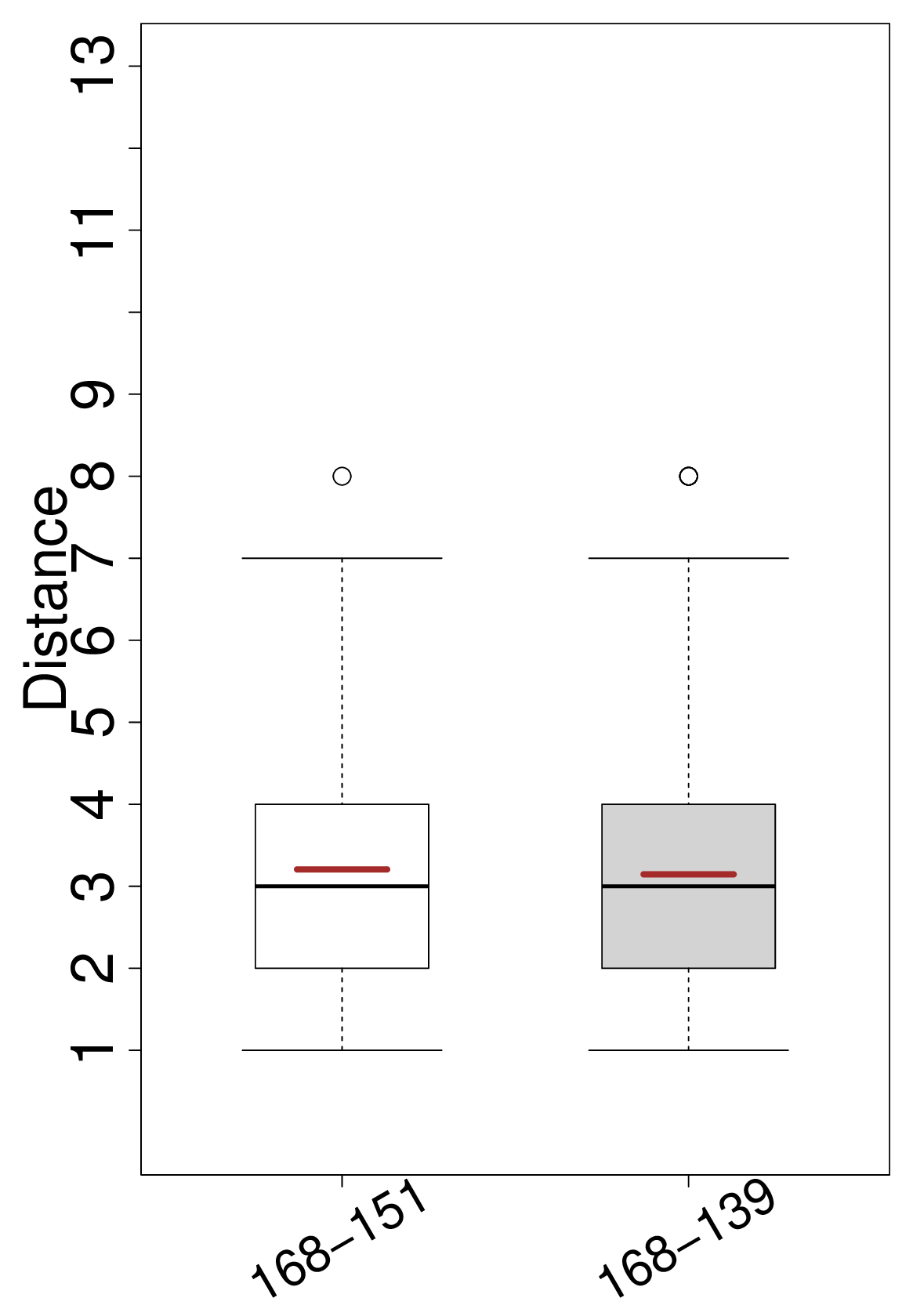}& 
\hspace{-0.5cm} \includegraphics [angle=-0,width=0.3\textwidth] {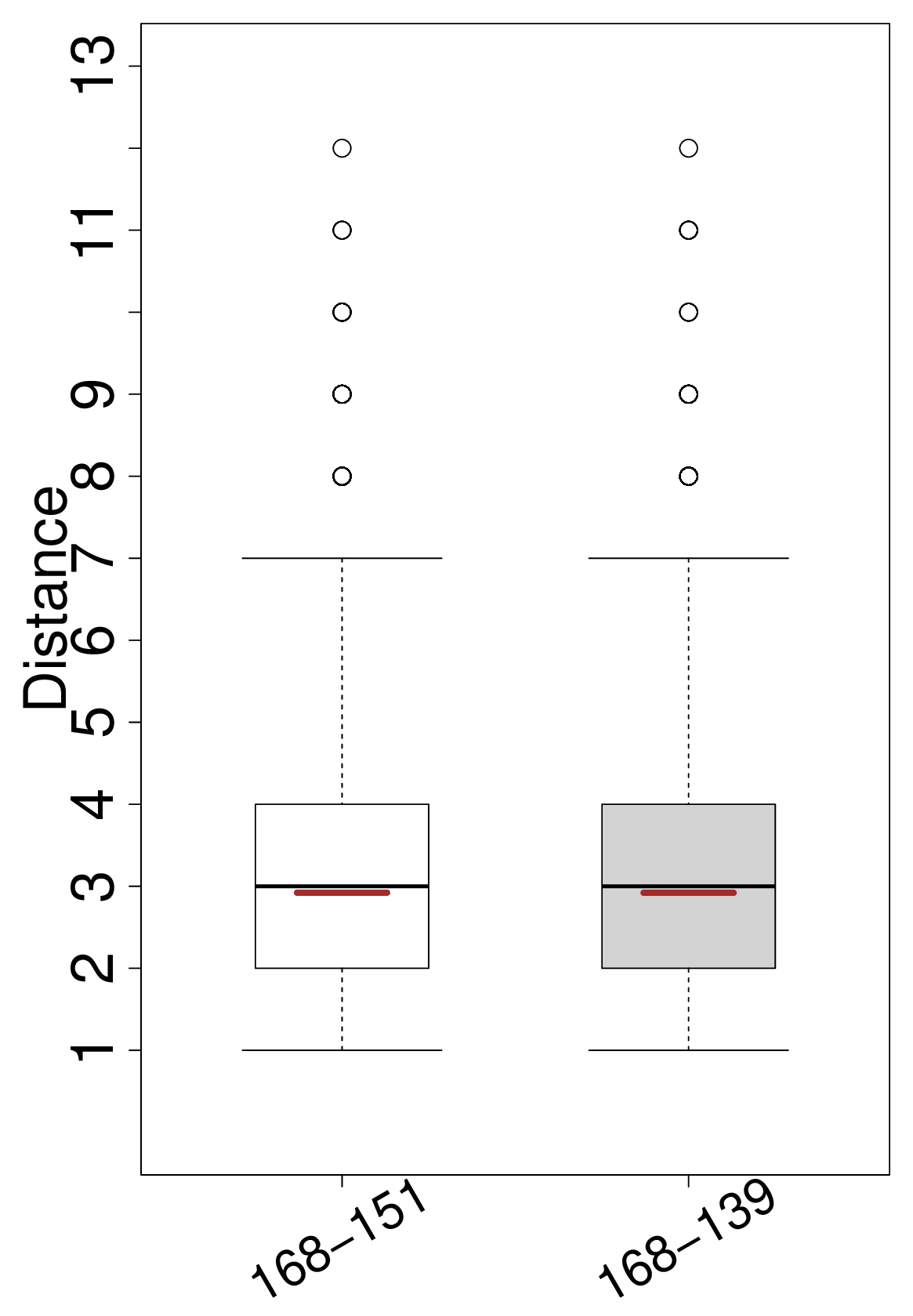}&
\hspace{-0.55cm} \includegraphics [angle=-0,width=0.3\textwidth] {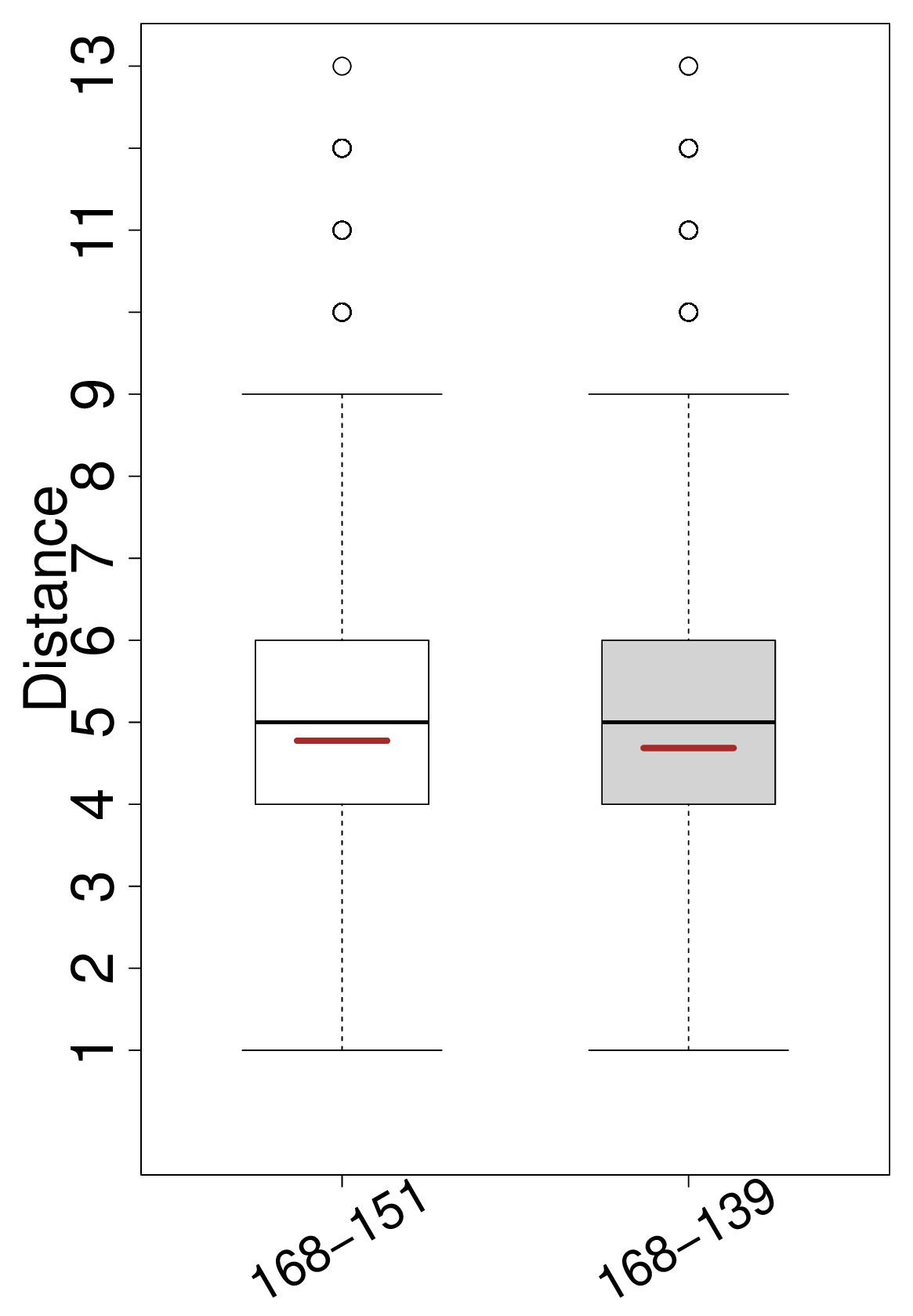}
\end{tabular}
\end{center}
\caption{Distribution of distances between terms $s\in C_i$ generating a new annotation path fork (with reference to term $k$) for protein $i\in V_{np}^k$, and the term $q$ where the fork is located. (\textit{yeast} (a), (b), (c), \textit{human} (d), (e), (f)). Red segments represent the distribution mean.
\label{fig:dist_level}}
\normalsize
\end{mdframed}
\end{figure}

Figure~\ref{fig:dist_level} reports the empirical distribution of distances between $q$ and $s$. We found that the maximum level observed in the GO is $16$, which accordingly is also the maximum possible distance between $s$ and $k$.
The median of distances between $q$ and $s$ span from $2$ (MF--yeast) to $5$ (BP--yeast, human), with the lowest values in the MF branch for yeast and in CC and MF  branches for human. The maximum distance in CC and MF is $7$, with few exceptions (outliers in the plot), whereas the minimum is $1$, meaning that the fork tend to be located relatively close to $s$, with at least 3/4 of cases within a distance $5$.
 On the other side, in the BP branch the fork tend to be sensibly more far from $s$ in the hierarchy: in both organisms the distance between $q$ and $s$  can be even $13$, meaning that terms $s$ and $k$ reside in two regions semantically less related or unrelated. Fortunately, this case represents a large minority, with 3/4 of forks being far from $s$ not more that $6$ edges.      
 BP is the most numerous and complex among the GO branches, and accordingly it does not surprise obtaining different trends on this branch. 
  
Overall, this experiment shown that previous and novel annotations for a protein tend to be located, at least in the half of cases, within a maximum distance $5$  from the common ancestor where an annotation path fork is formed. Nevertheless, we believe such a result needs to be further investigated, to better understand whether the outliers in different branches share common features. For instance, it would be interesting to assess whether forks much distant from $s$ in Figure~\ref{fig:dist_level} are to some extent related to low rank cases in Figure~\ref{fig:GO_rankings}, since distant terms in the ontology are expected to have a low similarity -- at least in terms of Lin similarity, which also takes into account the hierarchical relationships among terms.  

\section{Evaluation of negative selection}\label{sec:negsel}
By exploiting the results shown in Section~\ref{sec:analy}, we propose a novel negative sample selection algorithm, Negative Selection through Functional Similarity (NSFS), 
leveraging the semantic similarities among protein functions to discriminate whether that protein is a reliable negative example or not. Then, in order to provide a reference for the quality of our algorithm's negative examples, we carried out an experimental validation including the state-of-art heuristics used for negative example selection in the GO hierarchy. 
\subsection{Experimental setting}\label{exp}
We employed the temporal holdout setting for validating the quality of negative selection algorithms (see Section~\ref{prelim}): namely, algorithms infer negatives using the old release of annotation ($\bY$), and their predictions are evaluated on the later release ($\overline{\bY}$). The performance is measured in terms of number of false negatives averaged across terms, where a protein $i\in V$ is a \textit{false negative} for the term $k$ if the algorithm selected $i$ as negative for $k$ and $\overline Y_{ik} = 1$. GOA releases $69-52$ for yeast  and   $168-151$ for human are employed.  
To have a fair comparison, a negative selection method is given a \textit{budget} $B$ which is the number of negatives the method  must choose: if the method is not able to select enough negatives for a given term, the remaining negatives are selected randomly with uniform distribution (strategy \textit{Random}).
This setting is the same adopted in the benchmark evaluation~\citep{NOGO}.  
In order not to average across terms without proteins received novel annotations in the holdout period, we selected the terms $k$ for which $|V_{np}^k| > 0$, obtaining $96$ (resp. $383$), $150$ ($770$) and  $481$ ($2374$) CC, MF and BP terms in yeast (resp. human).
\subsection{\Name}\label{sub:snfn}
\begin{figure}[t]
\begin{mdframed}[style=mystyle2]
\begin{center}
\begin{tabular}{cc}
\hspace{-0.5cm} (a)  & \hspace{-0.3cm} (b)\\[-7pt]
\hspace{-0.5cm} \includegraphics [angle=-90,width=0.5\textwidth] {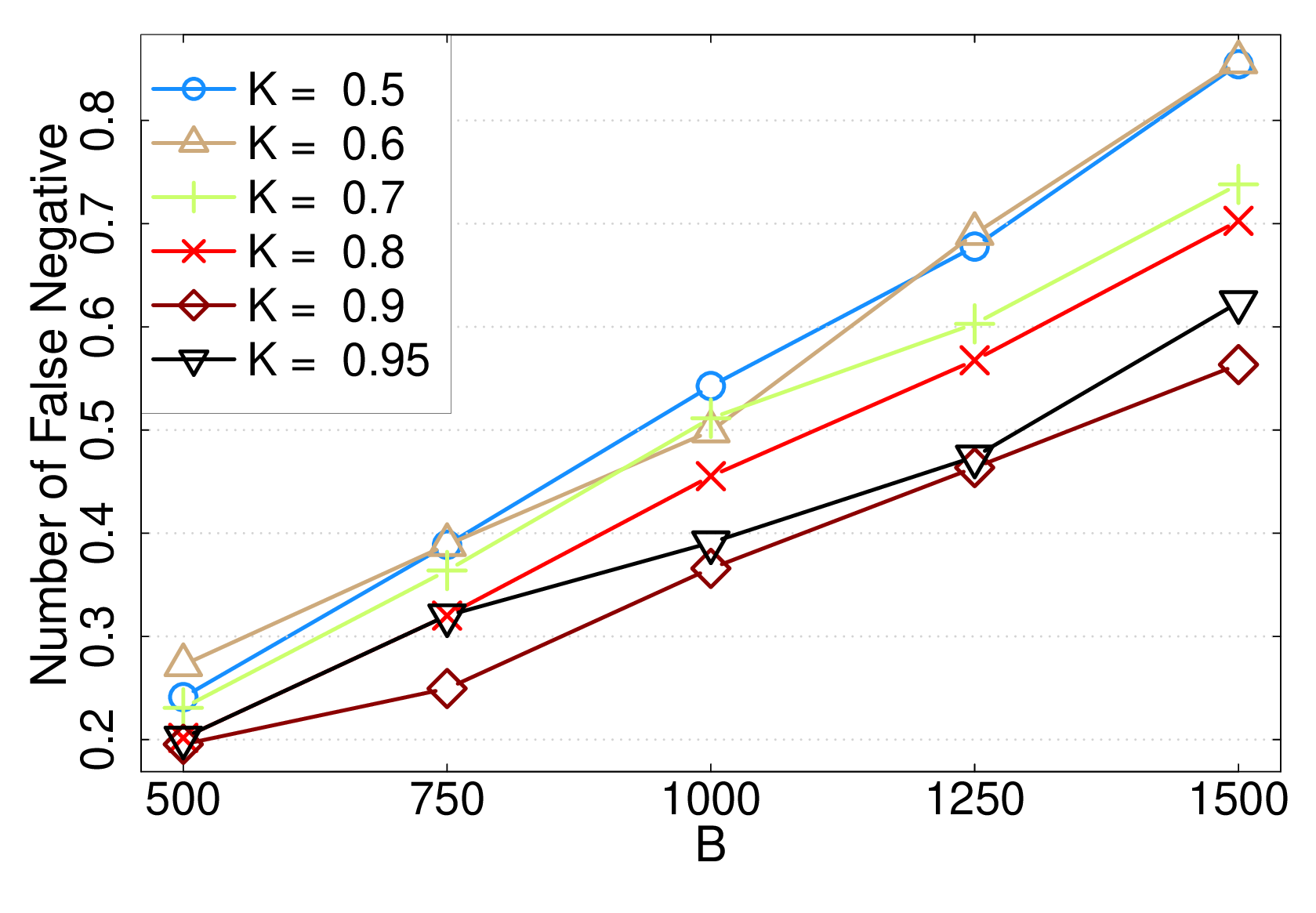}& 
\hspace{-0.3cm} \includegraphics [angle=-90,width=0.5\textwidth] {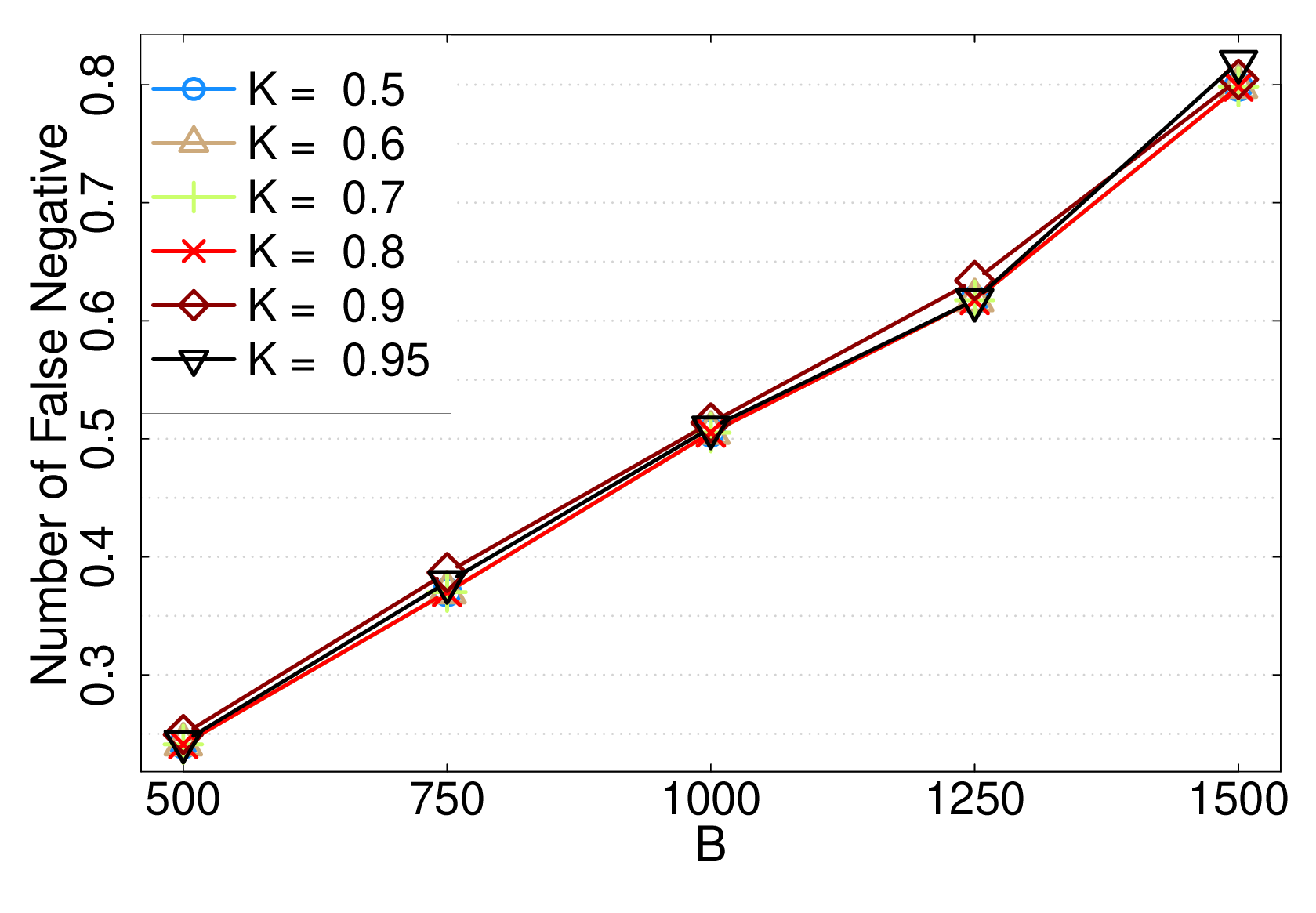}
\end{tabular}
\end{center}
\caption{Average number of false negatives of \Name~when tuning parameter $K$ on yeast data (BP, GOA releases $69$-$52$) using (a) Jaccard  and (b) Lin similarity measures.\label{fig:tuning}}
\end{mdframed}
\end{figure}

The  Negative  Selection through Functional Similarity (NSFS) heuristic relies upon the current annotation matrix  $\bY$ and the functional/semantic similarities $\bPhi_{k.}$ between the current function/term $k$ and the other available terms $C$. Recalling that $C_i=\{s \in C | Y_{is} =1\}$ is the set of terms a protein $i$ is already annotated with, and fixed a parameter $K \in (0,1)$,  \Name~selects as negative examples for term $k$ the proteins $i$ such that:
\begin{enumerate}[1.]	
\item $Y_{ik} = 0$,
\item $Y_{is} = 0$ for each $s\in C_i$ such that the similarity value $\phi_{ks}$ is above the $K$-th quantile of the similarity vector $\bPhi_{k.}$. Informally, $i$ is not annotated with terms most similar to $k$.   
\end{enumerate} 
This approach is motivated by the results shown in Section~\ref{sub:locating}, where we found that novel annotations for a given protein tend to fall on terms  very similar to the terms the protein was already annotated with.
\Name~thereby depends on the similarity matrix $\bPhi$, and consequently in the validation experiments we tested two variants: \Name-J and \Name-L, where $\bPhi$ stores the Jaccard ($\phi = \phi_2$) and Lin ($\phi = \phi_1$) similarities, respectively.
Furthermore, the method depends also on the parameter $K$, which we learn from the training data through internal cross validation. Nevertheless, to provide an idea of the impact this parameter has on the performance of the method, we supply in Figure~\ref{fig:tuning} the results of \Name~on yeast data and BP ontology when setting $K\in \{0.5, 0.6, 0.7, 0.8, 0.9, 0.95\}$. On the other data sets we obtained similar results (data not shown).
 
\Name~is more sensitive to $K$ when using the Jaccard instead of the Lin similarity, where the method obtains almost the same number of false negatives when $k$ varies. This is coherent with the differences in the distributions of similarities these two measure shown in Figure~\ref{fig:GO_rankings}, and with the fact that the ranks of terms $C_i$ are much closer when the Lin similarity is adopted (boxplots more compressed).
Results in Figure~\ref{fig:tuning} (a) instead show that the accuracy of \Name~tends to increase with $K$, except for $K=0.95$, where the method is likely to become too selective, and consequently the remaining negatives to reach the budget have to be chosen randomly. Clearly, the optimal $K$ is related to both $B$ and the number of proteins $|V|$ in the organism.
\subsection{State-of-the-art comparison}
We compared the following heuristics for negative selection proposed for the GO hierarchy: 
\begin{description}
\item \textit{Sibling}~\citep{Mostafavi09}. Negative examples for term $k$ are the proteins 
\[\{i\in V| Y_{ik} = 0 \wedge \exists~s\in\sib(k)~s.t.~Y_{is}=1\};
\]
\item \textit{NoAncDesc}~\citep{Eisner05}. Negative examples for term $k$ are the proteins $i$ with no annotation in any ancestor or descendant term of $k$, i.e.:
\[i\in \{V| Y_{ik} = 0 \wedge \forall s\in\{\anc(k)\cup\desc(k)\}~Y_{is}=0\};
\]
\item \textit{SNOB}~\citep{NOGO}. To compute negative examples for a given term $k$, this heuristic is based on the conditional probability $\hat p(k|r)$ of seeing an annotation for $k$ given an annotation for $r$. Each protein $i\in \{j\in V| Y_{jk}=0\}$ is associated with a score $\sigma_i = \frac{1}{|C_i|} \sum_{s\in C_i} \hat p(k|s)$, and proteins $i$ with the lowest score $\sigma_i$ are selected as negative examples. SNOB has been the top method in a recent evaluation of negative selection algorithms for GO functions~\citep{NOGO}.
\item \textit{Random}. The negative examples are uniformly selected from the set $\{i\in V| Y_{ik}=0\}$. This method is used as a baseline.
\end{description}
The budget has been set  to $B\in\{500, 750, 1000, 1250, 1500\}$, considering that the number of annotations (positives) for most terms is already lower than $500$, and  that machine learning algorithms need more balanced input data, since usually they show a sharp decline in their performance when input labeling are highly unbalanced toward negatives~\citep{Japkowicz02}.

\begin{figure}[t]
\begin{mdframed}[style=mystyle2]
\begin{center}
\begin{tabular}{ccc}
\hspace{-1.3cm} (a)  & \hspace{-2.7cm} (b) & \hspace{-2.7cm} (c)\\[-9pt]
\hspace{-1.3cm} \includegraphics [angle=-90,width=0.5\textwidth] {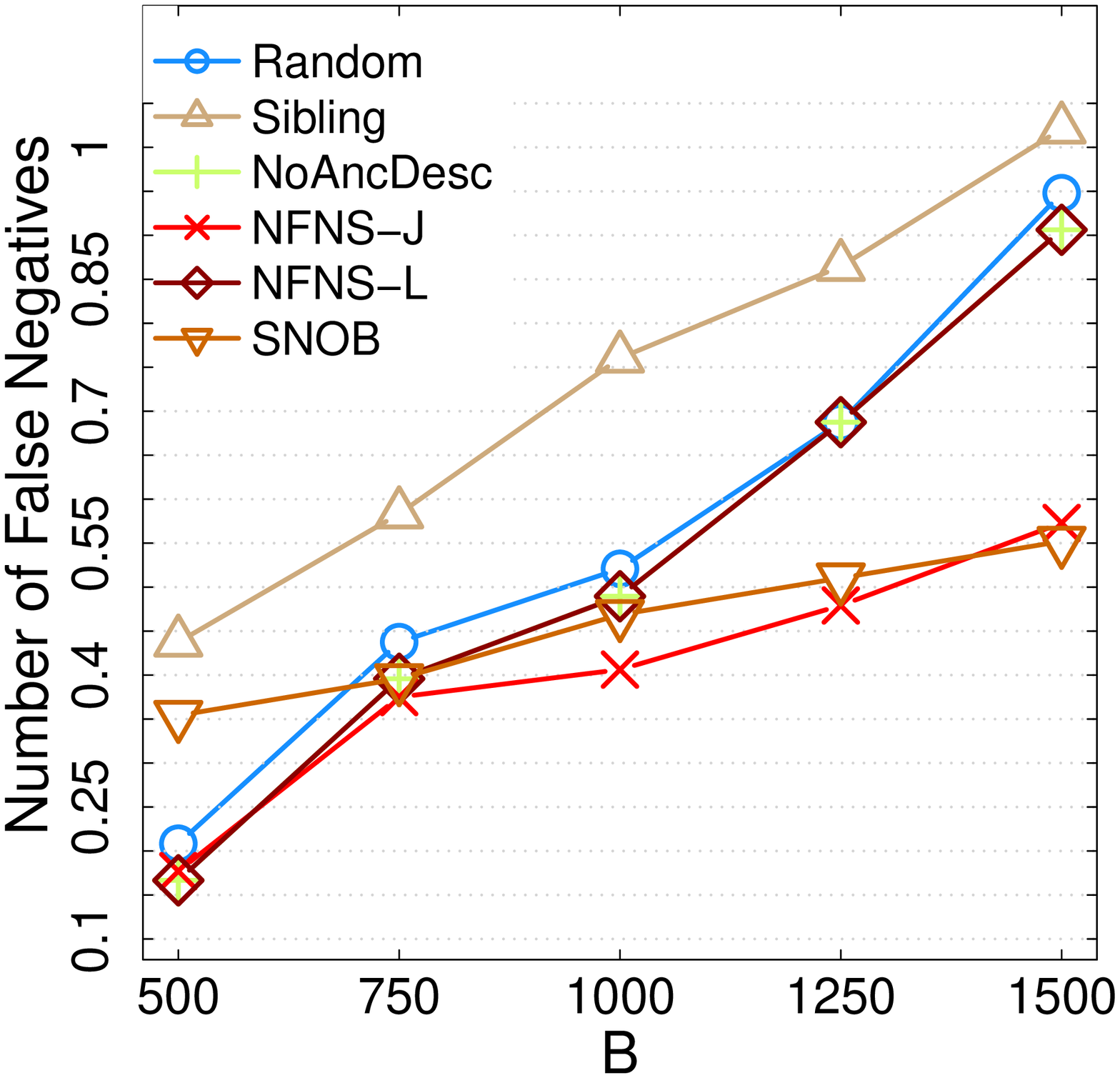}& 
\hspace{-2.7cm} \includegraphics [angle=-90,width=0.5\textwidth] {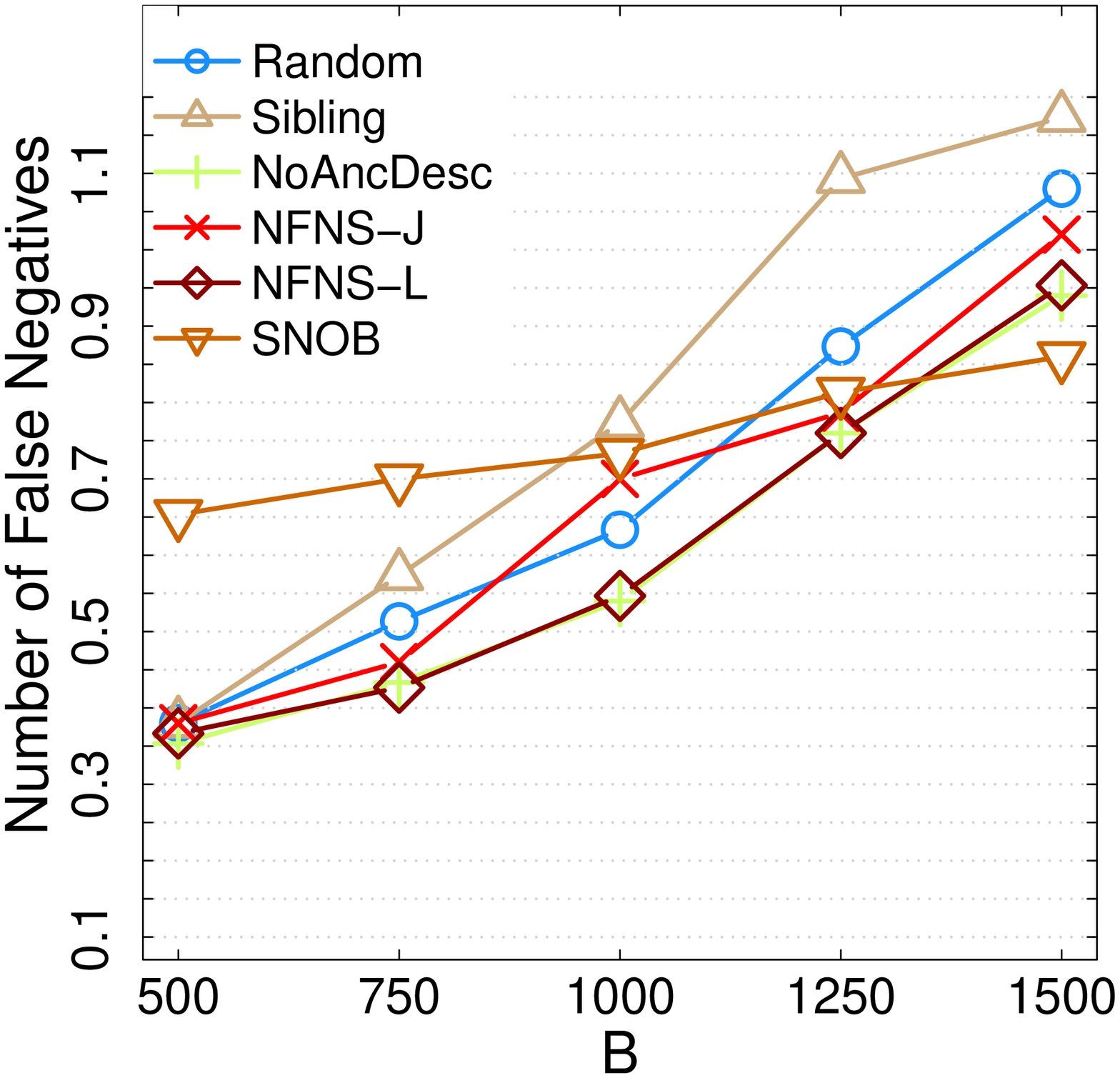}&
\hspace{-2.7cm} \includegraphics [angle=-90,width=0.5\textwidth] {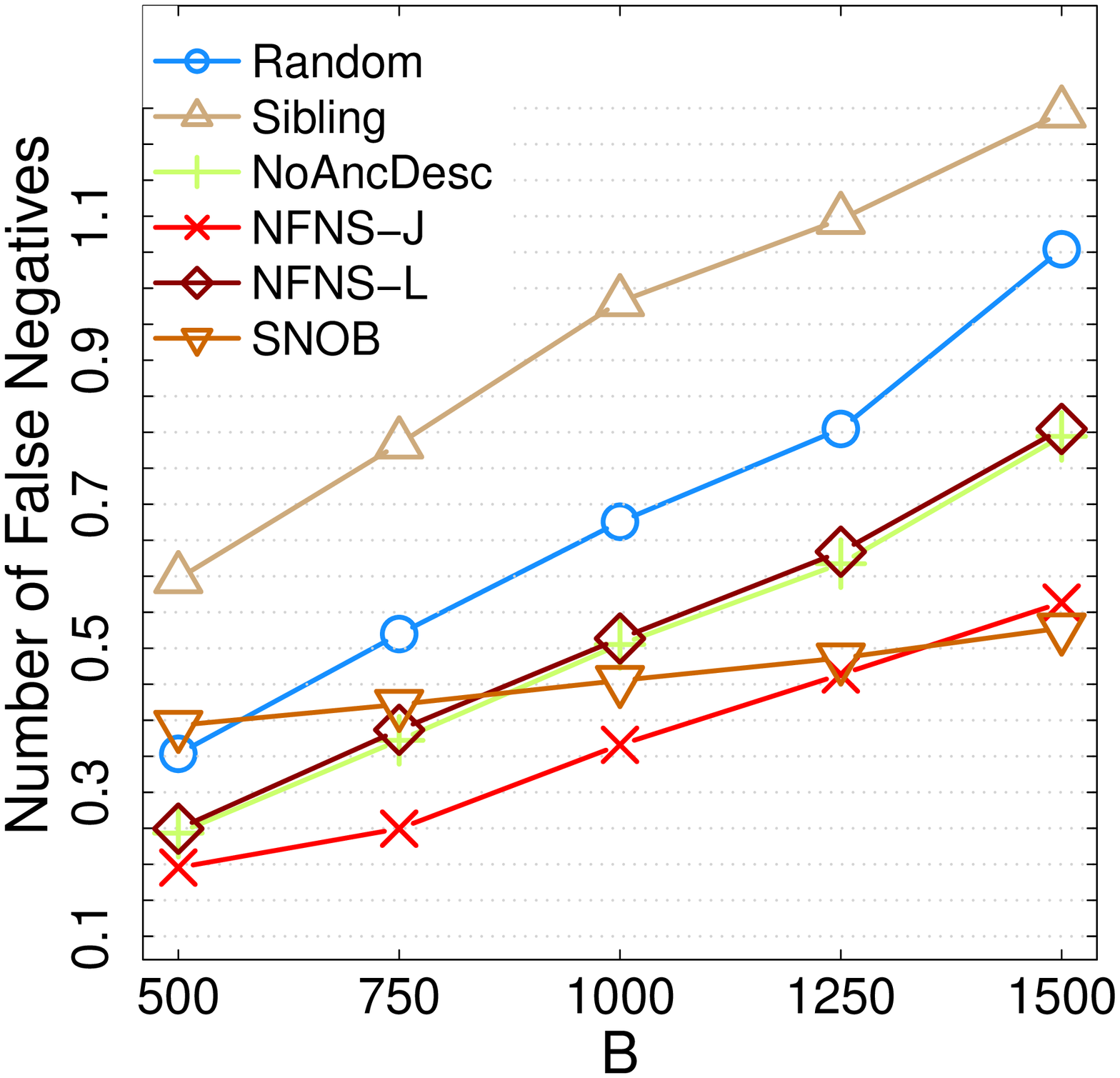}\\[2pt]
\hspace{-1.3cm} (d)  & \hspace{-2.7cm} (e) & \hspace{-2.7cm} (f)\\[-9pt]
\hspace{-1.3cm} \includegraphics [angle=-90,width=0.5\textwidth] {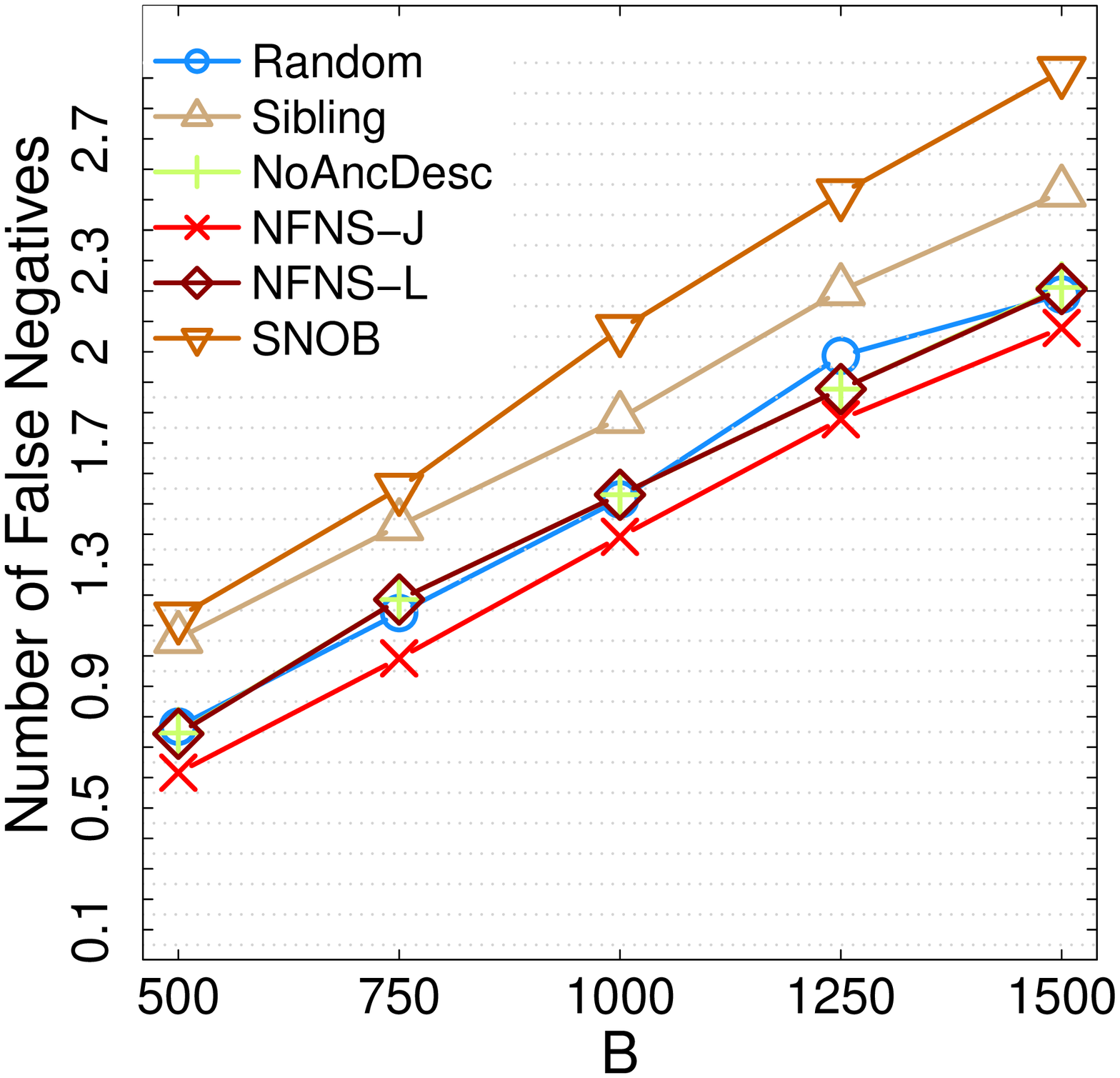}&
\hspace{-2.7cm} \includegraphics [angle=-90,width=0.5\textwidth] {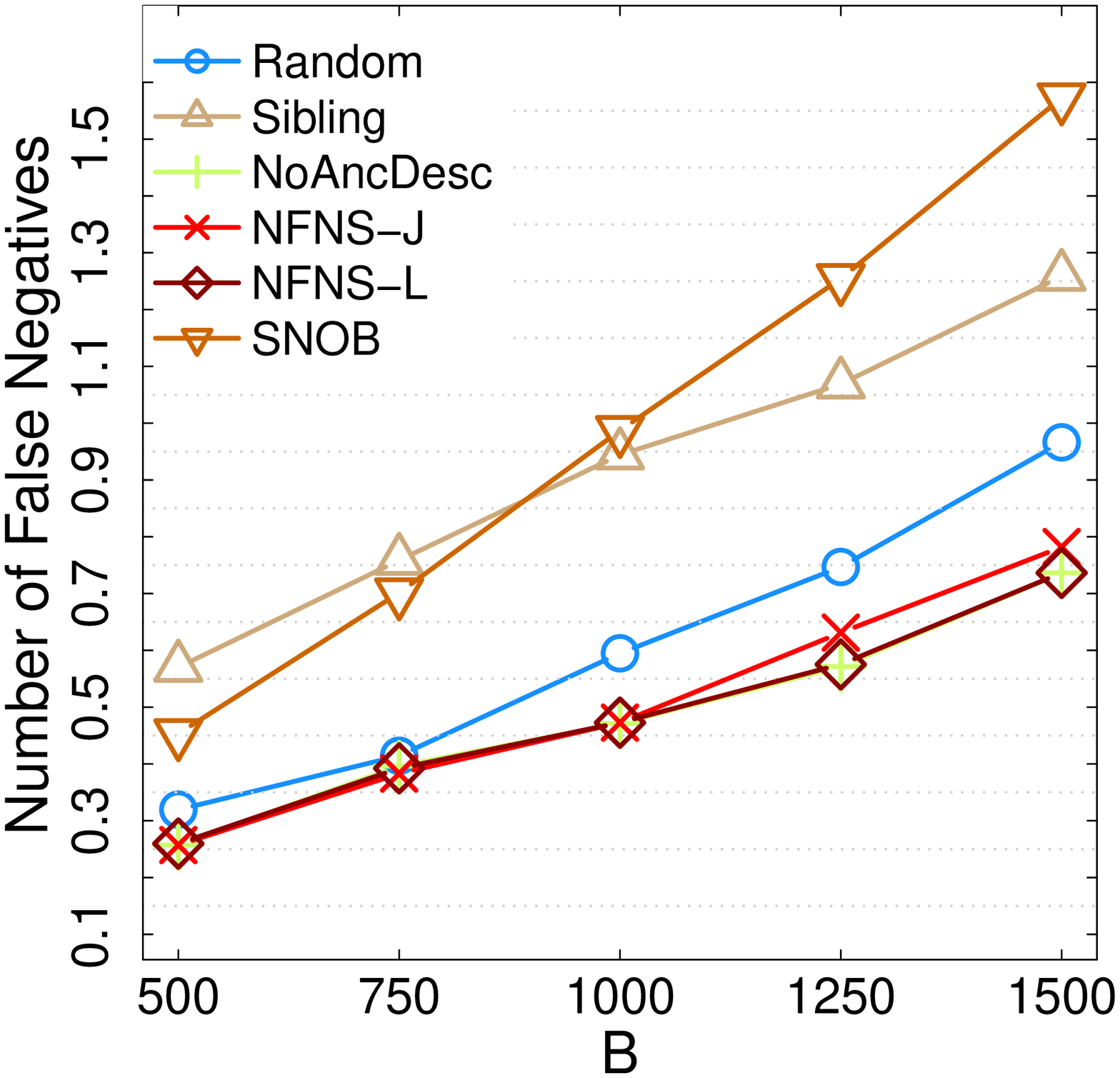}&
\hspace{-2.7cm} \includegraphics [angle=-90,width=0.5\textwidth] {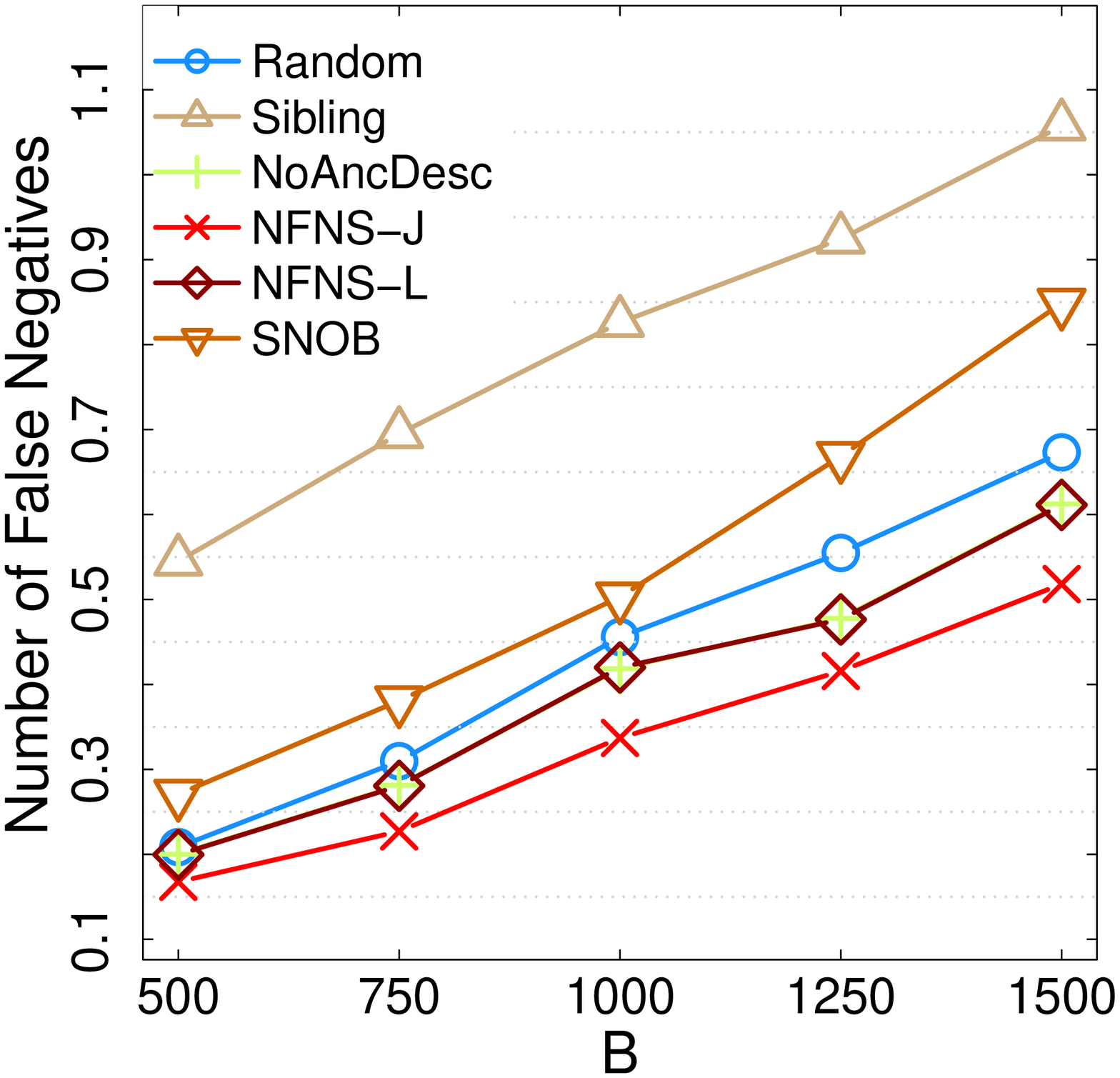}
\end{tabular}
\end{center}
\caption{Number of false negative averaged across GO terms for yeast (a)  CC, (b)  MF, (c)  BP) and human (d)  CC, (e)  MF, (f)  BP.\label{fig:false_negative_pred}}
\end{mdframed}
\end{figure}

The comparison results are presented in Figure~\ref{fig:false_negative_pred}. 
Sibling strategy poorly performs in all the experiments, confirming results found by \cite{NOGO}. The best performance is achieved by \Name-J on CC and BP branches in both yeast and human, with improvements statistically significant according to the Wilcoxon signed rank test, $p$-$value < 0.05$~\citep{Wilcoxon}.
On the MF branch, \Name-L performs slightly better than \Name-J, but with negligible and not statistically significant improvements. The SNOB heuristic achieves competitive results on yeast, mainly for higher budgets: when $B=1500$, it has the lowest average number of false negative predictions.  Nevertheless, on human data its performance dramatically declines, being in some settings still worse than sibling and random heuristics. This is quite in contrast with results proposed in~\citet{NOGO}, where SNOB always performed better than this two methods; this difference is likely due to the fact that authors in that study used also GO annotations with IEA evidence (inferred through electronic annotation), thus obtaining a set of enriched annotations  but more noisy. Indeed, author observed that the SNOB performance decayed on more specific terms, i.e. those having less annotations (information): thus removing the (even noisy) information coming from IEA annotations might cause the performance decline of SNOB.  Moreover, the subset of GO terms they adopted was different, namely the terms with $3$--$300$ annotations, which are partially included in our GO terms (those having at least one novel annotation in the holdout period).  
Finally, the NoAncDesc strategy performs similarly to \Name-L, always outperforming sibling and random heuristics, and with worse results that \Name-J on CC and BP branches.

Overall, these results confirm the insights provided by the analysis conducted in Section~\ref{sec:analy} about the distribution of reliable negative examples, and show that \Name~is able in effectively embedding them into a negative selection strategy.

\section*{Conclusion}
This work extensively investigated the evolution of protein annotations in different successive temporal releases of the Gene Ontology (GO) repository, with the aim of detecting reliable negative examples for automated algorithms that infer protein functions. We found that novel annotations for a given protein tend to appear on terms with high semantic similarity with the terms the protein was already annotated with in the previous GO release.  We experimentally verified this annotation trend on yeast and human organisms, and designed a novel method, \Name, leveraging it to effectively select negative examples. \Name~favorably compared with the state-of-the-art heuristics for negative selection in GO when selecting negatives on two organisms and for thousands of functions.

\section*{Disclosure statement}
 Authors declare no conflict of interest. 

\section*{Funding}
This work was supported by the grant title \textit{Machine learning algorithms to
handle label imbalance in biomedical taxonomies}, code PSR2017$\_$DIP$\_$010$\_$MFRAS,  Universit\`a degli Studi di Milano.

%

%
%
%
%

\end{document}